\documentclass[sigconf,nonacm]{acmart}

\usepackage{tikz}
\usepackage[nameinlink,capitalise,noabbrev]{cleveref}
\usepackage{hyperref}
\usepackage{microtype}
\usepackage{enumerate}
\usepackage{enumitem}
\usepackage{todonotes}
\usepackage{xfrac}
\usepackage{pifont}
\usepackage{adjustbox}

\newcounter{cccnt}
\newcounter{bdscnt}
\newcounter{rgcnt}
\newcounter{bccnt}

\renewenvironment{quote}{%
  \list{}{%
    \leftmargin0.5cm   
    \rightmargin0.05cm
  }
  \item\relax
}
{\endlist}

\AtBeginDocument{%
  }

\begin{document}


\title{A New Framework of Software Obfuscation Evaluation Criteria}

\author{Bjorn De Sutter}
\email{bjorn.desutter@ugent.be}
\orcid{0000-0003-0317-2089}
\affiliation{
\institution{Ghent University}
\city{Ghent}
\country{Belgium}
}



\begin{abstract}
In the domain of practical software protection against man-at-the-end attacks such as software reverse engineering and tampering, much of the scientific literature is plagued by the use of subpar methods to evaluate the protections' strength and even by the absence of such evaluations. Several criteria have been proposed in the past to assess the strength of protections, such as potency, resilience, stealth, and cost. We analyze their evolving definitions and uses. We formulate a number of critiques, from which we conclude that the existing definitions are unsatisfactory and need to be revised. We present a new framework of software protection evaluation criteria: relevance, effectiveness (or efficacy), robustness, concealment, stubbornness, sensitivity, predictability, and cost.
\end{abstract}



\maketitle


\newcommand{\defeq}{\overset{\mathrm{def}}{=}}

\section{Introduction}
\label{sec:introduction}

During the past three decades, many software obfuscation, tamperproofing, and anti-analysis techniques have been presented that aim to hinder reverse engineering and software tampering to prevent unauthorized use~\cite{collbergbook,survey2016,desutter2024evaluation}.

The progress in such software protections (SPs) has triggered an arms race in which many new techniques for deobfuscation, tampering, and analysis have been proposed. When used offensively, these techniques are collectively called \emph{man-at-the-end} (MATE) attacks, because the adversaries using them have full control over the systems on which they run, analyze, and alter the software under attack. Their tools include emulators, debuggers, disassemblers, custom operating systems, fuzzers, symbolic execution, sandboxes, and instrumentation among a wide range of static and dynamic analysis and tampering tools.
These give them white-box access to the software internals, including their internal execution state.

In the MATE attack model, and with the current state-of-the-art in practical SP, complete prevention of attacks is impossible. With enough time and resources to exploit their white-box access, adversaries can always succeed. The goal of MATE SP is hence to make the adversary's return on investment negative, by delaying the identification of attacks and by limiting their exploitation.

These goals are much fuzzier than the goals in some other security domains such as cryptography, which builds on precisely defined mathematical foundations to protect against man-in-the-middle attacks.\footnote{While cryptographic techniques and well-defined, cryptographically-sound concepts have also been developed in the domain of obfuscation, the results are still far from ready for widespread general use in practice. They are therefore outside the scope of this work, which focuses on practical, general-purpose software obfuscation.} Because of this fuzziness, and because MATE\linebreak attackers have so many alternative attack strategies and techniques at their disposal, it is difficult to measure the strength of SPs and to make founded statements about them.

As observed at the 2019 Dagstuhl Seminar on SP Decision Support and Evaluation Methodologies~\cite{Dagstuhl} and in a recent survey~\cite{desutter2024evaluation}, the field of MATE SP research suffers from a lack of standardized evaluation methodologies and practices. The survey observed that 
\begin{itemize}
    \item 25\%~$\hat{=}$~125/495 of the papers with SP implementations report no strength measurements at all.
    \item 54\%~$\hat{=}$~133/248 of the goodware obfuscation papers that deploy SPs on samples do not evaluate those SPs' strength by assessing how analysis methods fare against them.
    \item Obfuscation research often does not assess resilience.
    \item The costs of applying SPs are measured much more frequently than their strengths.
    \item Potency is measured much more often than resilience.
\end{itemize}

Although there has long been a common understanding that the relevant criteria for evaluating obfuscation techniques include potency, resilience, stealth, and cost~\cite{collberg1997taxonomy,collberg1998manufacturing}, there is no universally agreed and applicable method to define, qualify, or quantify them. 
Instead, non-compatible definitions and interpretations have been proposed and used in the literature. We believe that this lack of consistency contributes to today's unsatisfactory evaluation practices in MATE SP research and that the time has come to revisit the essential quality criteria of software obfuscation techniques.

In this paper, our aim is to study how the concepts of potency, resilience, and stealth have been considered in past research, how they have evolved over time, and where confusion about them may have arisen that may have been a contributing factor to the issues mentioned above. For doing so, we will pinpoint issues with the different definitions and interpretations of those concepts that we observed in past research, and we will sketch a new framework that can help to better frame those concepts, and provide advice as to how to consider them in future research. 

Sections 2--5 discuss four different approaches to define and interpret the concepts of potency, resilience, and stealth, and formulate critiques on them. Section 6 proposes an alternative framework of evaluation criteria. We provide further discussion in Section 7. Finally, Section 8 draws conclusions.

\section{Collberg, Thomborson, and Low}

In 1997 and 1998 Collberg, Thomborson, and Low (CTL) proposed three criteria to evaluate obfuscating transformations~\cite{collberg1997taxonomy,collberg1998manufacturing}.

\subsection{Definitions}

\subsubsection{Potency}
In 1997, CTL~\cite{collberg1997taxonomy} introduced the term potency to denote the degree to which an obfuscation confuses a \emph{human} reader trying to understand a given program, i.e., to quantify how the obfuscated program is more obscure, complex, or unreadable than the original program. They defined the potency that is obtained on a given program as
the ratio between a complexity metric computed on that program after and before the obfuscation~\cite{collberg1997taxonomy}.

\if false
follows~\cite{collberg1997taxonomy}:
\begin{quote}
  Let $\mathcal{T}$ be a behavior-conserving transformation such that $P \overset{\mathcal{T}}{\rightarrow} P'$ transforms a source program $P$ into a target program $P'$. Let $E(P)$ be the complexity of P, as defined by one of the metrics [...].

  $\mathcal{T}_{\text{pot}}(P)$, the \emph{potency} of $\mathcal{T}$ with respect to a program $P$, is a measure of the extent to which $\mathcal{T}$ changes the complexity of $P$. It is defined as
  $$\mathcal{T}_{\text{pot}}(P) \defeq E(P')/E(P)-1.$$ 
$\mathcal{T}$ is a \emph{potent obfuscating transformation} if $\mathcal{T}(P)>0$.
\end{quote}
\fi

They proposed considering software complexity metrics such as program length~\cite{halstead}, cyclomatic complexity~\cite{mccabe1976}, nesting complexity~\cite{harrison1981}, data flow complexity~\cite{oviedo80}, fan-in/out complexity~\cite{flow_metrics}, data structure complexity~\cite{munson1993}, and OO Metric~\cite{OO-metric}. CTL then derived and suggested that obfuscations should aim to directly impact these metrics, by increasing program size and introducing new classes and methods, by introducing new predicates and increasing nesting levels of conditional and looping constructs, etc.~\cite{collberg1997taxonomy}.

In line with the suggestion, various other complexity metrics have been used in SP research since then, such as QMOOD understandability~\cite{Foket14,qmood}. Various more advanced forms of complexity metrics have been proposed, some of which are specific for evaluating obfuscations, e.g., based on entropy~\cite{entropy_metric}, others more general, such as the cost of mental simulation~\cite{mental_metrics}.
Several authors have later revisited the idea of combining many different metrics~\cite{anckaert,D4.06}.

\subsubsection{Resilience}
\label{sec:resilience}
CTL observed that it is trivial to increase potency in ways that will not confuse a human reader, e.g., by injecting code that can immediately be identified as dead. They then presented the additional quality criterion of \emph{resilience} to differentiate useless transformations and useless impacts on complexity metrics from useful ones. In their proposal~\cite{collberg1997taxonomy}, resilience ``measures how well a transformation holds under attack from an automatic deobfuscator.'' 

It does so in terms of two measures. The \emph{programmer effort} is ``the amount of time required to construct an automatic deobfuscator that is able to effectively reduce [the obfuscation's] potency.'' The \emph{deobfuscator effort} is ``the execution time and space required by such an automatic deobfuscator to effectively reduce [the obfuscation's] potency.'' They scaled programmer effort in terms of the scope of the analyses and transformations underlying the deobfuscator, which can be local, global, interprocedural, or interprocess. The deobfuscator effort is based on the computational complexity of the underlying analysis: polynomial or exponential. Based on these properties, resilience is measured on a scale of \emph{trivial} to \emph{full}, plus the level \emph{one-way} that was based on the then realistic assumption that some information about the original program can never be recovered from the obfuscated program, such as identifiers names (of variables, classes, functions, etc.) and code layout.

\subsubsection{Stealth}
Although resilient transformations may not be susceptible to attacks by automated deobfuscation, they may still be susceptible to attacks by humans. In 1998, CTL~\cite{collberg1998manufacturing} hence added the \emph{stealth} criterion, which denoted ``How well does obfuscated code blend in with the original code?'' They initially did not formalize this criterion beyond stating that obfuscated code should resemble original code as much as possible, pointing out that stealth obviously is a highly context-sensitive metric. For example, they argued that graph-based opaque predicates will be stealthy in many Java programs as those often involve pointer-rich data structures. 

\subsubsection{Cost}
Finally, CTL argued that the impact of obfuscations on execution time and space should be assessed~\cite{collberg1997taxonomy}. Like stealth, \emph{cost} is a context-sensitive metric: injecting code into an inner loop or at the topmost program level will have a radically different impact. 

\subsection{Critique}
\label{sec:9798-critiques}
CTL's definitions of potency, resilience, and stealth do not satisfy. 

\subsubsection{Complexity Metrics Unfit for Obfuscated Code}
\label{sec:unfit_complexity_metrics}
The validity of software complexity metrics for code comprehension is questionable. Feitelson argues that most metrics are based on intuition instead of systematic empirical validation and that they fail to adequately quantify comprehension difficulty~\cite{feitelson}. The most popular metrics of code length and cyclometic complexity do not capture that the code reader's prior experience and background knowledge also contribute significantly~\cite{halstead,mccabe1976,code_metrics}. Other metrics such as cognitive functional size~\cite{cognitive_functional_size} do take into account how the human brain perceives different code structures and patterns, but only to a limited extent. If complexity metrics to model comprehension difficulty of regular software are already contested, expanding their use to assess the potency on obfuscated code is even more controversial. With their observation that the proposed metrics can trivially be increased with transformations that lack any utility as obfuscations, CTL themselves admit that complexity metrics cannot be reused as is for assessing the qualities of obfuscated code. 

When Foket et al.~\cite{Foket14} used validated complexity metrics for object-oriented software, they had to admit that their use was problematic. Additional qualitative analysis was needed to ensure that the quantitative metrics results were correctly interpreted and to avoid drawing invalid conclusions. 

Feitelson also assessed the importance of meaningful variable names for program comprehension~\cite{feitelson}, and found that their impact on comprehension difficulty varies from program to program. In obfuscated code, one should not expect to find meaningful variable names, at least not without using machine learning techniques to reconstruct them~\cite{2017_recovering_clear_natural_identifiers_from_obfuscated_js_names,banerjee2021variablerecoverydecompiledbinary}. This difference alone puts in doubt whether software complexity metrics developed on regular software can be reused as is for obfuscated code.

\subsubsection{Unclear How to Compute Complexity Metrics}
\label{sec:unclear_computations}
The definition of potency does not specify which software fragments to measure.  

First, the definition is vague about the program representation on which to compute the complexity metrics. Should one use ground-truth representations from the forward engineering process (e.g., source code, IR code used in a compiler, assembly code, binary code with full symbol and relocation information) or representations built during the reverse engineering process (e.g., IR code built by a disassembler, IR obtained by code lifting, or source code obtained after decompilation)? Which representations are more relevant: the ones being transformed to deploy the obfuscations, or the ones on which adversaries might get their hands?

For managed languages such as Java, this question might not be that important. The semantic gap between the Java source code and the distributed bytecode is relatively small~\cite{batchelder2007obfuscating} and a representation to compute metrics on (such as a call graph and method CFGs) can be built mechanistically from the distributed bytecode. 
For stripped binaries, the reconstruction of a suitable representation is much more complex. For example, when control flow and design obfuscations such as those by Van den Broeck et al.~\cite{jens21} are deployed, the difference between what the obfuscator knows to be the functions and what the disassembler thinks to be the possible functions is huge. Each disassembler, possibly even each version, configuration, and customization by means of scripts or plug-ins will yield different CFGs and different functions. In other words, there is no one representation on which attackers can get their hands. Even if all disassemblers would behave the same, do you compute the control complexity metrics on the assembly code, on the low-level IR lifted from the assembly, on the high-level IR built on top, or on the decompiled source code? These are open questions. Some work even skips disassembly and decompilation entirely, e.g., by interpreting executable files as bitstreams and then applying image processing machine learning techniques on those bitstreams to classify malware~\cite{ni2018malware,marastoni2021data} and to classify used obfuscations~\cite{secrypt21}.

Secondly, on what part of the software should the metrics be computed? Although adversaries might initially not know which part to study, they have many ways to zoom in on the most relevant parts, thus only requiring deeper study of parts of the program. But which parts exactly? And what about the parts they prune from their search space? Obfuscation can be useful in hindering that pruning~\cite{coppens2013feedback,reganoL2P}, rather than in hindering the understanding of the code that actually embeds the assets the adversaries are after. On which parts of the code should the metrics then be computed?

\subsubsection{Fuzzy and Confusing Boundary between Automated and Human Analysis}
\label{sec:fuzzy_boundary}
The definitions by CTL attempt to make a clear separation between automated analysis and deobfuscation on the one hand and human analysis on the other. We think this is not warranted. In practice, reverse engineers often make use of interactive tools that mix automated and human analysis, such as interactive disassemblers and decompilers of which the built-in heuristics can be overridden on demand by their users. Moreover, they always perform their manual tasks on data produced by automated tools: No one starts studying the bits of a binary executable manually. There is ample evidence for this in the literature~\cite{emse2019,Sutherland2006,Votipka2019,practice_malware,desutter2024evaluation}.

Moreover, by binding automation to deobfuscation in the term resilience and by binding human activities to confusion in potency, CTL beg the question of where to place automated attacks that do not deobfuscate code, such as automatic secret extraction attacks~\cite{khunt++,banescu15}. Banescu et al.\ use the term resilience to denote the impact of SPs on the time that symbolic execution engines need to extract a secret~\cite{banescu15,2017_predicting_the_resilience_of_obfuscated_code_against_symbolic_execution_attacks_via_machine_learning}, even though these engines make no attempt at producing deobfuscated code. This is a clear example of how hard it is to use CTL's terms resilience and potency consistently. 

\subsubsection{Tool Availability and Sharing over Time}
\label{sec:tool_availability}
The dependence of resilience on programmer effort neglects that once someone publishes an analysis tool or deobfuscator, everyone can reuse it. This certainly has an impact on the practical value of obfuscations. For example, in experiments with professional pen testers~\cite{emse2019}, it mattered a lot to them that QEMU was not able to correctly simulate binaries with anti-debugging~\cite{abrath2016tightly}, and that valgrind had to be adapted to support trace sizes beyond 2 GiB. 

Tool availability raises an interesting question for academics that can be illustrated with a well-known example: Intel's Pin instrumentation tool only works on the x86 architectures, with no similar tool being available for ARM these days. Then does that mean that dynamic analysis of ARM code is fundamentally harder than dynamic analysis of x86? For an academic studying reverse engineering, that seems intuitively wrong, even though, for a practicing reverse engineer, it may be actually true.

\subsubsection{Narrow Definition of Programmer Effort}
Using only one dimension to estimate the effort required to program an analysis is flawed. In addition to the scope, at least the sensitivity~\cite{path-sensitive} and polyvariance~\cite{polyvariance} of the analysis should also be considered.  However, it is not clear whether more precise analyses are always more difficult to implement than less precise ones. It is hence not clear how to rank different forms of precision in terms of programmer effort. Then again, if a framework were available to instantiate various forms (i.e., different scopes and different sensitivities) of a custom analysis with a simple configuration, the programmer effort would no longer be dependent on them. 
   
\subsubsection{Narrow Scope of Deobfuscation}
Resilience was defined in terms of deobfuscation, meaning ``returning the obfuscated code into a form that is as easily understandable as the original source''. Pure deobfuscation is rarely used in practice. Instead, adversaries often try to bypass SPs and work around them~\cite{emse2019}.

Moreover, the local, global, and interprocedural programmer effort levels proposed by CTL~\cite{collberg1997taxonomy} are strongly biased toward static analyses and static program representations such as call graphs and CFGs, as opposed to dynamic representations such as traces that are also used for dynamic deobfuscation~\cite{generic_deobfuscation,mariano24}. Should we hence not use other aspects to rate the effort required to obtain dynamic information with a certain precision? Some dynamic approaches, such as the bit-level tracking by Yadegari et al.~\cite{bitleveltaint}, are clearly more complex to implement and execute than the simpler taint-tracking approaches proposed by Faingnaert et al.~\cite{khunt++} or the alternative from Li et al.~\cite{khunt}, so it seems that a more flexible scheme is necessary to rate deobfuscation techniques.

\subsubsection{Asset and Attack Goal Independence}
The definitions of potency and resilience are asset independent. They qualify and quantify how confusing obfuscations are, in general, for human readers, and how susceptible they are to an adversary's countermeasures. Their formulation does not take into account that adversaries aim to violate specific security requirements on specific assets, such as the confidentiality or integrity of specific code fragments. 

The literature is clear on the fact that for assessing the strength of obfuscations, both the adversary's goal and the deployed software analysis methods need to be considered. Schrittwieser et al.\ distinguish four different goals: finding the location of data, finding the location of program functionality, extraction of code fragments, and understanding the program~\cite{survey2016}. The latter can be interpreted broadly. For example, determining invariants of a program, determining preconditions for the execution of some fragment, or obtaining concise summaries of its functionality can all be considered program understanding goals. Schrittwieser et al.\ also distinguish four classes of software analysis methods that obfuscations might aim to mitigate: pattern matching, automatic static analysis, automatic dynamic analysis, and human analysis. While one could argue that different software metrics should be used depending on the type of assets to be protected and on the considered attack goals and used analysis methods, as has been done in the past~\cite{anckaert}, it is all but clear which complexity metrics to use when.

Furthermore, the proposed code artificiality metric for stealth only is applicable in the face of specific attacks, namely those considering n-grams or related syntactical code properties, but meaningless for other code and data location finding attacks. 

\subsubsection{Layering is Missing in Action}
\label{sec:layering_missing}
Nowhere do CTL's definitions stress the need to evaluate the marginal strength of obfuscations when they are composed and layered on top of each other, despite the fact that layering obfuscations is the best practice~\cite{layering,collbergbook,recipes}.

Because the obfuscation (singular) being evaluated could consist of a composition of multiple transformations (plural), the definitions by CTL support layering and compositions and are technically fine. 
However, to help the community, we do think that the concepts of layering and composability need to be put forward explicitly, and potential pitfalls or consequences deserve to be discussed, if only because layering increases the relevance of other critiques, such as those discussed in Section~\ref{sec:unclear_computations}. 
When only one obfuscation deployed in isolation is being evaluated, the semantic gap between the different representations on which defenders and attackers can compute the relevant metrics might be small enough that we do not need to worry about which representation to compute it. But once obfuscations are composed, the semantic gap can grow quickly and the question of how to compute the metrics becomes much more difficult again. De Sutter et al.\ discuss this in more detail where they observe the lack of layered SP evaluation in research~\cite{desutter2024evaluation}.

\subsubsection{Machine Learning Changed the Nature of the Game}
The concept of one-way resilience no longer applies in this day and age of machine learning as it did in the previous millennium. Even if machine learning techniques offer no guarantee of reconstructing the original identifier names or the original code layout, they have been proven capable of reconstructing meaningful, useful names and layouts that help human reverse engineers~\cite{2017_recovering_clear_natural_identifiers_from_obfuscated_js_names,banerjee2021variablerecoverydecompiledbinary}. However, one-way resilience might still be meaningful when obfuscations rely on one-way functions, such as hash functions~\cite{hash1,hash2}.

\subsubsection{Lack of Sensitivity}
\label{sec:sensitivity_missing}
CTL neglect the sensitivity of potency, resilience, and stealth with respect to features of the program being obfuscated. An example of an obfuscation that can be highly sensitive to (local) program features is the source-level injection of mixed Boolean arithmetic expressions that compute constant values~\cite{mba}. If the variables used in the expressions happen to have constant values, or are uninitialized, compilers can optimize away (part of) the obfuscation.\footnote{
Source code gives compilers much richer information than adversaries get from a binary. The fact that a compiler can deduce some property hence does not imply an adversary can do so, i.e., that the obfuscation fails to hide the property for an adversary.} Another example is bogus control flow and bogus code inserted by means of, e.g., opaque predicates. The effect on complexity there depends entirely on which data dependencies and which control dependencies happen to be impacted by the bogus code. In general, combining obfuscation with program optimization is known to be a difficult problem~\cite{optimization}.

This sensitivity is an issue for at least two practical reasons. First, if the impact of an obfuscation is sensitive to features of the program being obfuscated, a mechanism is needed to assess that impact every time the obfuscation is considered as a candidate for deployment. Predictive techniques or actual measurements can be used for this. The latter is not really feasible, however, given the large design space that needs to be explored~\cite{Basile23}, and the former has significant limitations today~\cite{reganoMetric}. Secondly, programs evolve over time. So even when users of an obfuscation tool have somehow obtained a suitable composition of obfuscations for their software v1.0, if that composition is sensitive to program features, they will need to reassess its suitability for later versions. Clearly, the user-friendliness of obfuscators and/or their need to include predictive capabilities depends on the sensitivity of the supported obfuscations with respect to program features. 

CTL also neglect the sensitivity of an obfuscation to its configuration parameters. If an obfuscation's impact is sensitive to them, this again impacts the user-friendliness of obfuscators and the need to include prediction mechanisms, because the parameters will need to be optimized. One could argue that if two configurations of some obfuscation have significantly different impacts, then they should be treated as two different obfuscations. Although documenting obfuscations as such may make an obfuscation tool more transparent, it does not make the obfuscation selection and optimization problems any easier. For these reasons, the sensitivity of obfuscations should be considered a prime quality criterion. 

\subsubsection{Learnability Neglected}
\label{sec:learnability_missing}
Defining stealth as the artificiality of transformed code compared to original code neglects that adversaries can learn to single out obfuscated fragments even if they are similar to original code. If an obfuscation is implemented with exactly the same instruction sequence every time it is deployed, it does not matter that this sequence also occurs a few times in the original code: Pattern matchers will quickly achieve 100\% recall.

\section{Nagra and Collberg}
In 2009, Nagra and Collberg (NC) redefined potency and stealth~\cite{collbergbook}.

\subsection{Updated Definitions}
\label{sec:collberg_nagra_def}
NC now defined potency in relation to a program's property that some adversary might try to reveal, and in relation to analyses that can be used to reveal that property. First, they paraphrased the meaning of an \emph{effective obfuscating transformation} as one that ``makes it harder to perform the necessary analyses that reveal the secret property on the obfuscated program than it is on the original.''

Given a specific program, a specific property thereof, and some specific analysis that can reveal that property on the original program, an obfuscation is considered \emph{effective} if the analysis can no longer reveal an equivalent property on the obfuscated program or requires more resources to do so. The obfuscation is considered \emph{ineffective} if the analysis can still reveal an equivalent property using the same amount of resources and \emph{defective} if the analysis can reveal an equivalent property using less resources. 

In their definition, the term ``equivalent'' $\approx$ was not defined formally. Although in many cases an adversary aims for revealing a property exactly, NC argued that there can also be cases in which an adversary is happy with an approximation. Scenarios in which approximation may suffice include data or code localization attacks, in which it can suffice if a tool prunes most of the search space that the attacker has to explore manually, or cryptographic key extraction, in which it can suffice if an analysis tool can severely prune the search space for a brute-force attack. 

The analysis considered in the definition need not necessarily be a single analysis. It can in fact be a sequence of analyses and (deobfuscating) transformations; such as when analyses reveal properties that are useful to deobfuscate the code, and the property ultimately targeted by the adversary is the deobfuscated code. 

Based on the above definition, NC then defined \emph{potent obfuscating transformations} in terms of a specific program, a specific property, and a \emph{set of analyses}. An obfuscation is \emph{potent} if it is effective against at least one of the analyses and not defective against any of them: ``a potent obfuscating transformation makes at least one analysis harder to perform, and no analyses easier.''

NC omitted resilience as a separate, complementary quality criterion. The reason is, of course, that the above redefinition of potency already incorporates the original concept of resilience.

For stealth, NC distinguish between \emph{local stealth} and \emph{steganographic stealth}~\cite{collbergbook}. A transformation is steganographically stealthy if an adversary's detector function cannot determine if a program has been transformed with it or not. A transformation is locally stealthy if the adversary's locator function cannot tell the locations where the transformation has been applied.

Formally, they propose to define the local/steganographic stealth of an obfuscation with respect to a considered class of programs and a considered locator/detector function as the maximum of the expected false positive rate and the expected false negative rate of that function over all programs in the class after obfuscation. 

With respect to detector and locator functions, there exists quite a bit of research on automatic techniques for the classification of software as obfuscated versus unobfuscated and for the automatic determination of the deployed obfuscations and obfuscators. Such techniques are invariably based on the observation of software features that form fingerprints of obfuscations whose presence can be considered as indicators of (a lack of) stealth~\cite{obfuscation_detection}.

Schrittwieser et al.~\cite{modeling_stealth} used complexity metrics as fingerprints. Raubitzek et al.~\cite{obfuscation_detection4} convert binary code bytes to grayscale values and use singular value decomposition to uncover patterns created by different obfuscation techniques in images. Jiang et al.~\cite{obfuscation_detection2} use the number of different types of instructions in basic blocks and a basic block adjacency matrix of function control flow graphs as features, as well as the number of string constants in basic blocks. Bacci et al.~\cite{obfuscation_detection3} and Wang and Rountev~\cite{obfuscation_detection5} use strings and bytecode n-grams as features. Kim et al.~\cite{obfuscation_detection6} use opcode distribution. Tofighi-Shirazi et al.~\cite{obfuscation_detection7} obtain symbolic expressions from disassembled function bodies through symbolic execution and then use the words in them and their frequencies as features in a bag-of-words approach. Tian et al.~\cite{obfuscation_recognition} represent functions by their so-called reduced shortest paths. Zhao et al.~\cite{obfuscation_detection8} use assembly instruction embeddings based on skip-grams, convolutional neural networks (CNNs) to generate basic block embeddings from the instruction embeddings, and long-short-term memory networks (LSTM) to encode the semantics of basic blocks and their relations into features. Kanzaki, Monden, and Collberg proposed a code artificiality metric based on an N-gram instructions model~\cite{code_artificiality}.

These features are all used in classifiers, of which the models serve to quantify stealth in the sense that an obfuscation can be considered more stealthy if the models perform more poorly on it. 

\subsection{Critique}
\label{sec:2009-critiques}
The redefinition of potency answered many of the critiques from Section~\ref{sec:9798-critiques}. However, it still lacks in numerous ways. Some critiques still apply, such as those on layering (\ref{sec:layering_missing}), sensitivity (\ref{sec:sensitivity_missing}), and learnability (\ref{sec:learnability_missing}). Additional critiques are the following. 

\subsubsection{All Analyses Treated Equally}
\label{sec:all_equal}
In the definition of a potent obfuscation, the characteristics of the analyses against which it must be effective and must not be defective are not considered. Suppose that an obfuscation is defective with respect to one analysis because it decreases its computation time on a program from 10s to 5s, while being effective for another analysis because it increases that analysis' computation time from 2s to 8s. If the adversary has both analyses available, their minimal execution time increased from 2s to 5s with the obfuscation; their combined time increased from 12s to 13s. So should that be considered a potent obfuscating transformation? According to the current definition, it is not. 

The new definition of potent transformations implicitly treats all considered analyses equally important to mitigate, as if attackers have an oracle at their disposal that tells them which of all potentially useful analyses they should use given a concrete program they need to analyze. This worst-case scenario assumption can definitely be relevant in a risk management approach~\cite{Basile23} to ensure that potentially relevant attack strategies are not overlooked. However, for evaluating the practical strength of obfuscations, one clearly needs to consider which analyses adversaries are more likely to deploy. 
To some extent, this relates to stealth: When a deployed obfuscation is not stealthy, this might enable attackers to select the most appropriate analyses, or even to customize them. 

One might argue that one can counter this critique by selecting not a set of individual program analysis techniques to evaluate potency, but a set of multi-step attack strategies, in which each strategy consists of pre-pass analysis that is first deployed to select the most appropriate actual analysis technique to go after the targeted property, after which that technique is then used. However, this reformulation would only shift the problem from the actual analysis techniques to the pre-pass analyses, rather than solve it.

In computer security, in general, a clear distinction is made between security assessments and risk assessments. The former focus solely on the technical identification of vulnerabilities and exploits, while the latter also considers the likelihood that attackers know about an existing vulnerability and corresponding exploit, the likelihood that the attackers are willing to use that exploit, and the potential damage that such exploitation would have on the victim and their business. The proposed definitions of potent transformations and stealth provide no guidance on how to incorporate or evaluate the likelihood that adversaries will choose custom analyses, or how difficult it is for them to do so, i.e., how attackers can decide on the most fit-for-purpose analysis.

\subsubsection{Comparing Obfuscations}
One shortcoming of NC's definition of potency is that it does not allow one to compare different obfuscating transformations, e.g., to determine which one is more potent for some program, against which analyses. They in fact acknowledge that, and refer to the work by Dalla Preda et al.\ for a solution~\cite{mila07}. Section ~\ref{sec:abstract_interpretation} will focus on this solution.

\subsubsection{Ad Hoc Formulas}
NC's formulas for computing stealth compute the maximum of false positive and false negative rates. We do not see why this would be advisable over using more standardized metrics such as recall, precision, or the F1 metric. Many authors seem to agree with us. In the post-2009 literature on obfuscation detection techniques, including a 2016 paper co-authored by Collberg~\cite{code_artificiality}, not a single paper uses the formulas proposed by NC. 

\subsubsection{Resilience Missing in Action}
With the redefinition of potency, the original distinction between potency and resilience has become moot. Their unification definitely offers advantages, such as moving away from the artificial distinction between confusing humans performing manual work and confusing tools that automate analyses (and deobfuscating transformations). However, it also comes with major disadvantages. As observed by De Sutter et al., many publications in the domain of software obfuscations feature sub-par evaluations~\cite{desutter2024evaluation}. Particularly relevant for this work are their observations that few papers that present novel obfuscation techniques evaluate those obfuscations against real-world attacks and against adversaries that adapt their attack strategies to the fact that certain obfuscations have been deployed, as is commonly the case in the cat-and-mouse game of SP, and as has been observed in human experiments by H\"ansch et al.~\cite{2018_programming_experience_might_not_help_in_comprehending_obfuscated_source_code_efficiently}.

De Sutter et al.\ advocate to differentiate between potency and resilience because authors of obfuscation papers are recommended to consider at least two classes of attacks against which to evaluate the strength of their obfuscation techniques. In their vision, attacks on the assets being protected with some obfuscation determine its potency, and attacks on the SP itself determine its resilience. The distinction between potency and resilience is then still pretty vague, but the continued usage of the two terms can help to raise awareness about the best practices of SP evaluation methodologies.

\subsubsection{Confusion about Potency and Resilience}
These updated definitions have not been picked up by many colleagues. Instead, the original definitions of potency and resilience are still often used, and sometimes inconsistently, as we mentioned in Section~\ref{sec:fuzzy_boundary}. Perhaps this should come as no surprise: In 2017, 8 years after the redefinition of potency that abandons resilience in his book~\cite{collbergbook}, Collberg himself co-authored the work of Banescu et al.~\cite{2017_predicting_the_resilience_of_obfuscated_code_against_symbolic_execution_attacks_via_machine_learning} that still uses resilience when assessing the capability of symbolic execution to extract secrets instead of deobfuscating code.

\subsubsection{Layering Neglected, of Protections and Analyses}
Layering of SPs, and hence assessing the marginal strength of SPs, is still neglected. So is the combination of analyses. In practice, attackers rarely rely on a single analysis method or a single attack step. The definitions of NC do not reflect this. One could argue that it is the responsibility of the individual researchers to consider analyses that are combinations of code analysis techniques rather than individual ones. We believe that anyone defining the aspects to be evaluated should stress the usefulness of this and discuss, in general, the options to do so. 

\subsubsection{Narrow Scope of Stealth}
\label{sec:narrow_stealth}
The updated definition of stealth based on the accuracy of detector and locator functions instead of on the similarity between transformed code and original application code is certainly an improvement. When the results of locator and detector functions are merely used as data inputs to later attack steps, measuring their accuracy on obfuscated code can suffice to evaluate the obfuscation's stealth.

When their results are used as control inputs, i.e., to make strategic decisions on which attack steps to apply next or on how to configure the next attack steps, measuring their accuracy is still useful. In such cases, however, one has to raise the question if there might be other ways to enable the same decisions. For example, maybe an adversary only needs to know which obfuscation tool has been used to make a strategic decision. Or maybe they need to know a specific configuration parameter with which an obfuscation was applied in some location to make a strategic decision? 

By relying only on obfuscation locator and detector functions, one cannot assess to what extent other fingerprints of deployed protections can be revealed that enable strategic decisions. 

\section{Dalla Preda and Giacobazzi}
\label{sec:abstract_interpretation}
Dalla Preda and Giacobazzi, with a variety of collaborators, have worked extensively on theoretical foundations to make obfuscations comparable, namely through the lens of \emph{abstract interpretation} (AI). In their view, attackers are limited in what they want to analyze and in their resources, so they model them as a specific abstraction. The attacker then coincides with the considered analyzer.

Discussing all the formal foundations of their work and presenting their formal definitions is not feasible in this paper, as it would require too long an introduction to abstract interpretation and theoretical results achieved in that domain. Unlike the definitions discussed in previous sections, their results in the form of mathematical proofs and equations are not intuitive and hard to interpret for non-experts.  We therefore summarize their main conclusions and how their work has evolved in an informal manner.

\subsection{Definitions based on Completeness}
\label{sec:completeness}
Initially, the study of AI to define potency~\cite{cousot1977abstract,cousot2021principles} was based on the idea that the statement ``obfuscation makes programs incomprehensible for observers'' can be rephrased as ``obfuscation makes programs incomplete for abstract interpreters''~\cite{2008hiding}. The potency is therefore defined in terms of the \emph{completeness} of AIs with which attackers might try to reveal properties of a program that the defender is trying to prevent with obfuscations. 

An analysis of a program modeled as an AI is incomplete if the analysis produces an imprecise result. An imprecise result occurs \emph{when an analysis produces some valid result that is less precise than another valid result that the analysis can represent}. For example, consider a value set analysis that computes and propagates intervals of the form $[a,b]$ for any real values of $a$ and $b$ with $a \leq b$. If such an analysis reports that some variable in a program can hold values in the range $[0,10]$, while it can actually only hold values in the range $[1,5]$, the analysis has proven to be incomplete for that program. Importantly, incompleteness depends on both the program and the analysis. Some concrete analysis, and variations thereof, can be complete on one program and incomplete on another one. 

The mentioned authors define the potency of an obfuscation in terms of its impact on completeness: Which analyses become incomplete on which programs by deploying an obfuscation? 

Analyses modeled as AIs operate on abstract domains. The domains themselves consist of lattices, but all possible domains also form a lattice, meaning that they are ordered in terms of the relative precision of the properties that they can represent and potentially deduce from programs. And hence so are the analyses: Some can represent more precise properties than others, some can deduce more precise properties than others, and hence some analyses can deduce certain properties on more programs than other analyses.  

In older work~\cite{mila05a,mila05b,mila07}, an obfuscation is considered potent when there is a property that is not preserved, i.e., when some AI that could compute that property on the original program can no longer compute it after  obfuscation. Different obfuscations can then be compared based on the most precise properties they preserve. 

In 2012--2017, potency was redefined in terms of specific analyses that are used to reveal certain properties~\cite{2012_making,2017GMDP}. The potency of an obfuscation with respect to a specific analysis, program, and property is then determined by whether or not the completeness of the analysis on the program is impacted. If the analysis can no longer compute the property on the obfuscated program, the obfuscation is considered to be potent. Importantly, even if the analysis can still extract the property, the obfuscation might still be useful. Although the obfuscation then cannot counter that specific analysis, it might be able to counter simpler variants thereof. To assess this utility, the \emph{potency range} of an obfuscation was defined in terms of its impact on the completeness of compressed versions of the analysis, with ``compressed'' meaning ``less precise''.

In 2018, Bruni at al.~\cite{bruni2018code1,bruni2018code2} proposed an alternative method to assess obfuscations that aim to counter model checking attacks based on abstraction refinement, such as CEGAR~\cite{cegar}. In such attacks, a model checker iteratively tries to check the validity of models, starting with a very abstract domain that can lead to a conclusion quickly. When a conclusive result cannot be reached with the domain used in one iteration, a more precise refined domain is chosen, and a new attempt is made in a next iteration. The strength of an obfuscation is then defined as the extent to which it can prevent the checker from coming to a valid conclusion in fast early rounds with very AIs. Interestingly, the authors illustrate that many existing data flow analyses, such as liveness analysis, constant propagation, and available expression analysis, can be reformulated as model-checking problems, thus widening the applicability of this method. 

In follow-up research in 2022, Bruni et al.~\cite{bruni2022repair} present AI repair strategies to automatically refine domains to make the AI (locally) complete for a given program, with the goal of enabling program verification methods to start with any abstract domain. If the initial domain is too abstract to avoid false alarms, it can be refined with this strategy to produce complete results. The user of an AI then no longer needs to choose the domain beforehand to yield optimal formal verification results. In our eyes, this comes close to how reverse engineers adapt their strategies and analyses and how they move on to more complex ones if their initial, basic attempts fail. 

In 2022--2023, Campion et al.\ introduce the notion of $\epsilon$-partial incompleteness~\cite{Campion2022,campion23}. An abstract interpreter can be $\epsilon$-partial incomplete with respect to a given program and a given (set of) input values, meaning that the imprecision of the AI result is bounded by $\epsilon$, that is, the distance between the results of the abstraction of the concrete semantics (e.g., $[1,5]$ in the value set analysis example above) and the result of the AI on the given input (e.g., $[0,10]$) is at most $\epsilon$. For this, they define distance metrics on the abstract domains. They argue that the ability to quantify the amount of imprecision induced in the AI by an obfuscating transformation could be used to measure the potency of such a transformation.

Interestingly, they concede that, in general, one cannot automate the procedure of deciding whether the AI of a given program on a given input satisfies a given precision bound $\epsilon$. 
Still, the notions of $\epsilon$-partial incompleteness and bounded distance between actual properties and obtained analysis results can also open opportunities to reason about the strength of analyses that only have to compute approximate results (see Section~\ref{sec:collberg_nagra_def}). To the best of our knowledge, this direction has not yet been explored. 

In 2025, Giacobazzi and Ranzato~\cite{2025roberto} showed that the program property of having the best possible AI is not trivial and, in general, hard to achieve. Among others, they showed the impossibility of achieving the best correct abstraction property through minimal abstraction refinements or simplifications of the abstract domain. This puts into question the underlying assumption of some of the discussed definitions of potency that are based on which analysis can still reveal which properties. 

A strength of several works is that they not only allow to reason about the strength of obfuscations, but also include techniques to automatically derive obfuscations to counter the analyses defined in terms of AI~\cite{mila05a,mila05b,mila07,DP2013,DP2018,Roberto2012}. This includes obfuscations that target control flow analysis, data flow analysis, model checking, and more.

\subsection{Definitions based on Adequacy}
\label{sec:adequacy}
In 2023, Giacobazzi et al.~\cite{fitting_roberto} proposed to complement (and even replace) completeness as the basis for assessing potency with \emph{adequacy}. They observe that ``completeness characterizations do not really deal with the loss of precision due to the choice of the abstract observation, since they characterize only whether there is an extra loss of precision due to the computation on observed/abstracted data (compared with the observation of the concretely computed result).'' In other words, completeness is a measure of the mismatch between the computations in a given program and the chosen abstract domain (lattice) in which the program is interpreted, rather than measuring how good that domain is for revealing the properties in which an attacker/analyst is actually interested. 

Consider the trivial $\top$ abstraction that abstracts all concrete values to $\top$. Its lattice consists of one element $\top$. The result of such an AI on any program produces $\top$, which is by construction the most precise outcome that can be presented in the lattice, so such an AI is by construction complete. But it is completely useless: its result comes down to ``I don't know.'' In other words, the produced result is the least precise result that the attacker/analyst can be interested in, despite the interpretation being complete. 

Notice that such an element $\top$ is indispensable in many analyses: For an undecidable analysis to be sound, it needs to be capable of responding ``I don't know'', for which the used lattice includes $\top$.

The adequacy of an AI for a given program is then defined as the ability of the interpretation to produce analysis results for that program that contain strictly more information than $\top$. The relative adequacy of an AI for a program, i.e., adequacy with respect to some other element $\tau$ of the lattice, is then defined as the ability to produce a result that is strictly more precise than $\tau$.

The link with potency then is that an obfuscation can be considered potent with respect to some analysis and some program if it can make the analysis become inadequate on the obfuscated program while it was adequate on the original program. 

\subsection{Critique}

\subsubsection{Practical Applicabillity}
\label{sec:ai_practical}
While the use of AI theory to evaluate the strength of practical obfuscations against practical attacks has been discussed, such as obfuscations to mitigate disassemblers and slicing~\cite{2017GMDP}, it remains mostly a theoretical subject, of which the practical applicability is unclear. Although the AI work for the evaluation of model checking by Bruni et al.~\cite{bruni2018code1,bruni2018code2} might have the potential for wider applicability, a recent 571-paper literature review on evaluation methodologies in SP research~\cite{desutter2024evaluation} observes that software obfuscation and deobfuscation researchers do not have a strong appetite for this form of analysis: Model checking was one of the least popular attack methods, used in only 4\% of the papers that used concrete attack methods to evaluate obfuscations. 

\subsubsection{Attacks Success/Failure instead of Delay}
In the field of practical MATE SP, it is understood that MATE attacks cannot be completely prevented. Given enough time and effort, adversaries will always be able to get what they want. The main goals of obfuscation therefore are to delay attacks, to increase their costs, and to decrease their return-on-investment. In contrast, the existing research on AI and obfuscation focuses by and large on evaluating whether or not some property can be revealed with some interpretation/analysis and whether this can be prevented with an obfuscation. In other words, this research focuses on scenarios in which the defender tries to make attacks fail rather than trying to delay them. Although the complexity of different AIs is compared in some works to compare the strength of different obfuscations, the link between actual attack delay and analysis complexity is not made. This research hence seems to miss the point of MATE SP. 

\subsubsection{Soundness}
Research on AI and obfuscation only considers sound analyses. This is fine in many program analysis scenarios, such as program verification. In the MATE attack model, by contrast, attackers use any analysis, sound or unsound, that allows them to reach their goal. All dynamic analyses, e.g., are unsound. In this regard, this research again completely misses the point. 

\subsubsection{Program Understanding Only}
In 2008, Giacobazzi~\cite{2008hiding} stated ``The lack of completeness of the observer is therefore the corresponding of its poor understanding of program semantics''. This pinpoints an important limitation: This research, at least in the first order, targets software comprehension. It neglects the other MATE attack goals on obfuscated software. In addition to \emph{code comprehension}, Schrittwieser et al.~\cite{survey2016} identified \emph{finding the location of data}, \emph{finding the location of program functionality}, and \emph{extraction of code fragments} as important attacker targets. NC also claim that an adversary targeting an obfuscated program typically goes through a locate-alter-test cycle~\cite{collbergbook}. Neglecting location finding attacks is a clear limitation of AI research into the meaning of potency.

\subsubsection{Defender vs.\ Attacker Perspective}
\label{sec:perspective}
In much if not all of the cited research, the perspective of theory development is that of defenders that know which property of which program fragments they want to hide with obfuscations. The defenders hence know which analyses to consider for their assessment of the obfuscations' strength. Attackers, on the contrary, often do not know a priori which property they are after. In particular in data and functionality location finding attacks, the attackers by definition do not know the fragments of which they want to reveal properties.

Moreover, as defenders know which properties they want to hide, they can reason about the possible refinements of analyses that undo the impact of the chosen obfuscations. Attackers that try to find code or data or that try to understand a program do not know a priori which properties they are after. So how can they deploy similar refinement strategies? If they cannot, then why should defenders worry about the capabilities of refinement strategies? 

For most of its history, research on AI and obfuscation did not address the question of how difficult it is for an attacker to choose the best or simplest suitable AI for their attack. Only in 2025, Giacobazzi et al.\ started to address this issue, conceding that this may indeed be difficult~\cite{2025roberto}. This puts in doubt the relevance of using completeness or adequacy of the best possible abstractions for evaluating obfuscation strength. In our eyes, the defender should care mostly, if not only, about the analyses that an attacker will possibly or likely use, rather than on theoretically better, but in practice unlikely to be used, attack methods. 

An example might make this critique concrete. Consider interprocedural constant propagation~\cite{interp_cp}, and a scenario in which a context-insensitive variant of the analysis is not precise enough to reveal some interesting program property, while some $k$-depth context-sensitive variant is precise enough for some $k$. Undoubtedly, the context-sensitive variant, which will be slower and have a larger memory footprint, should be considered the most complex. There exist countless variations of the analysis with complexities in between those two variants, namely variations that analyze only specific parts of the program in a context-sensitive manner while handling the rest in a context-insensitive way. Intuitively, the variant that requires the least context-sensitivity should be considered the simplest one. Now, while the defender might know where context-sensitivity is useful and where not, hence being able to determine the simplest complete analysis for their program at hand, how is an attacker supposed to determine this variant? In particular when an attacker does not yet know the relevant properties to be revealed, this seems inconceivable. So what is the relevance of that best and simplest complete variant? 

One might think that attackers would go with the simplest or fastest analysis first and switch to more precise and slower ones only when the initial ones fail. So, first run, e.g., a context-insensitive analysis across the entire program (which should be cheap), then use whatever information they extracted to target a context-sensitive analysis to the most likely location of the program.

This is probably often how attackers work, but certainly not always. A counterexample is a crypto key extraction attack, for which Ceccato et al.\ performed experiments with professional pen testers~\cite{emse2019}. The most experienced among them skipped static analysis entirely, assuming that the deployed SPs would make that too hard, so they used only dynamic techniques from the start. In other words, they opted for a more complex analysis method over a simpler one, even though a simpler one would have worked perfectly well on an unobfuscated program.

In any case, even if the attackers do start with the simplest analyses and refine them as they obtain additional information, we think it is relevant to consider the effort required to extract that information and how precise it would need be, e.g., to pinpoint the most likely location in the program where more precision is needed. We should not assume that the refinement comes for free.

On this topic, there exists quite some work on on-demand data flow analysis or demand-drive data flow analysis. Such analyses are designed to be more efficient than blindly running the most complex analysis variants on a whole program. Those on-demand analyses, which adapt on-the-fly are more complex than analyses for which an a priori determination was made about which locations in the program require a higher form of sensitivity. So the consideration of such analyses does not solve the issue that AI research starts too much from the defender's perspective. 

The work by Bruni et al.\ cited above~\cite{bruni2022repair} is highly related. In that work, it is also assumed that the user will run a simple analysis first, and then iteratively run ever more refined versions thereof. All of that effort should be considered when evaluating potency or resilience, however, not just the complexity of the final analysis version. Maybe the journey is not more important than the destination, but it definitely is as important.

\section{Attack Tool Metrics}

A practical approach to evaluate the potency of obfuscations is to measure their impact on actual attack tools. One can compare how much resources a tool requires on vanilla and obfuscated programs, and how well the results obtained on the vanilla programs approximate those on the original ones. 

\subsection{Example Usage in Literature}
\label{sec:attack_tool_examples}
Examples are Foket et al.~\cite{Foket14} comparing points-to set sizes computed by WALA~\cite{WALA} on obfuscated Java programs, Linn and Debray~\cite{Linn2003} counting how many instructions no longer get disassembled correctly by objdump~\cite{binutils} and IDA Pro~\cite{IDA}, Van den Broeck et al.~\cite{jens21} counting how many CFG edges are no longer drawn correctly in CFGs displayed by the IDA Pro disassembler~\cite{IDA} and how many times code is duplicated in CFGs reconstructed with Binary Ninja~\cite{ninja}, and Cozza et al.~\cite{mila24} counting how many invariants among PLC (progammable logic controller) register values can be uncovered by Daikon~\cite{Daikon}. When such metrics can be categorized as true and false positives and negatives, quality measures such as recall, precision, accuracy, F1-score, etc.\ are often used~\cite{statistics}.

Papers that contribute deobfuscation techniques (e.g.,~\cite{2022_chosen_instruction_attack_against_commercial_code_virtualization_obfuscators,2021_mba_blast_unveiling_and_simplifying_mixed_boolean_arithmetic_obfuscation,2017_syntia_synthesizing_the_semantics_of_obfuscated_code,2015_a_generic_approach_to_automatic_deobfuscation_of_executable_code}) obviously evaluate the resilience of obfuscations. They do so by reporting the aforementioned quality measures for their contributed techniques on sample sets, and by comparing the similarity of the deobfuscated code to the obfuscated code. The same holds for papers that contribute obfuscation techniques and evaluate them against known deobfuscation methods (e.g.,~\cite{2022_loki_hardening_code_obfuscation_against_automated_attacks,2021_search_based_local_black_box_deobfuscation_understand_improve_and_mitigate}).

In their work on obfuscation strategies for industrial control systems, Cozza et al.~\cite{mila24} use two metrics to quantify resilience along two dimensions. First, they consider the number of invariants establishing connections between genuine and spurious registers. Genuine PLC registers are the ones that actually control physical processes, while spurious registers are the ones that seem to be used for such control (because of the obfuscation) but are not. The idea is that such invariants complicate the attacker’s ability to discern between those two types of registers and their true and fake usage. The second metric is the number of aggregations (clusters) of PLC registers associated with physical devices belonging to the same physical process. These aggregations supposedly make it difficult for the attacker to separate genuine from spurious physical processes handled by the same PLC. Cozza et al.\ use Daikon~\cite{Daikon} to extract the invariants from their use case applications.

\subsection{Critique}

\subsubsection{Ad hoc Nature}
\label{sec:adhoc}
Most tool-based metrics are ad hoc. 
For example, Van den Broeck et al.\ used two entirely different metrics for evaluating the impact of the same obfuscations on two disassemblers, because those two tools differ in the way in which they reconstruct functions and their CFGs~\cite{jens21}. For IDA Pro, which assigns each basic blocks to exactly one function, they report how true and fake control flow transfers are correctly interpreted and displayed in the CFGs. For Binary Ninja, which assigns each basic block to every function from which the block is reachable through intraprocedural control flow transfer idioms, they report the amount of resulting code duplication in the created CFGs.

Because of their ad hoc nature, it is questionable whether the conclusions drawn from the measurements are more generally valid, i.e., beyond the specific tools and tool versions that have been evaluated. Such answers can only be answered when sufficient insight into the internal operation of the tools used is available, either on the side of the authors of the publications or on the side of their readers. The latter requires that the papers contain sufficient information. This is definitely not always the case~\cite{desutter2024evaluation}.

Importantly, not only is the nature of the tool-based metrics often ad hoc, so is the deployment and configuration of the tools. For example, Van den Broeck et al.\ evaluated the use of IDA Pro with a number of simple plugins that helped the tool to overcome some of its most basic shortcomings resulting from the fact that it was not developed to handle the forms of code produced by the evaluated obfuscation~\cite{jens21}. In other words, Van den Broeck et al.\ evaluated the use of the tool in one of the possible ways in which attackers might use it after adapting their mode of operation to the existence of the novel obfuscation. This is a good practice~\cite{desutter2024evaluation}. Their plugins only covered a tiny fraction of the adaptations that an attacker might make, however, so how generally applicable are their results? Again, this question needs to be raised. 

A standardized reverse engineering analysis toolbox, with a commonly accepted set of analysis tools and commonly accepted measurement criteria might help to overcome the potentially limited validity of ad hoc metrics. At least it could lead to some standardization of evaluation methods, in particular when also the benchmark programs would be standardized. As observed in literature, however, no standardization is currently forming~\cite{desutter2024evaluation}. 

Attempts have recently been made to create a reusable and expandable toolbox and models to evaluate the strength of SPs~\cite{checkmate24}. It is unclear whether such a toolbox could ever be considered complete. For one, we do not think that there exists such a thing as a complete tool: attackers can use unsound techniques, and some techniques work better in some cases than others. To remain sound or soundish, many different heuristics and techniques can be used by tools. To optimize user-friendliness, additional heuristics and techniques can be used. All of these work better in some scenarios than in others, and importantly, they may not all be compatible. The already cited work Van den Broeck et al.\ is exemplary~\cite{jens21}. Because the policies of adding basic blocks to functions differ so much among the most popular disassemblers, some metrics simply are irrelevant and/or meaningless for some of them. 

The problem goes deeper even, because it is hard to come up with metrics that are relevant to as many scenarios as possible. For example, if a disassembler does not reconstruct functions correctly, how does one measure the quality of that reconstruction? After having given this question much thought, we did not find a good answer. To some extent, this issue is the same as with theoretic complexity metrics: They may be relevant and validated in one scenario (unobfuscated code), but this does not imply their relevance and validity in other scenarios (obfuscated code). 

A standard evaluation tool suite will certainly benefit our domain. The Common Criteria for certifying smart cards is a good example of similar standardization~\cite{common_criteria}. For the more philosophical/theoretical question of how to define potency, which is a non-functional feature, this will suffice, however. First, as discussed at several points already, attackers refine their strategies as they go: how to take into account the effort/difficulty/success of the most likely refinement strategies? Secondly, not all possible customizations/refinements of analyses can be included? How to take that into account? All in all, we do not think a complete tool is possible, and hence the definition or framing of the evaluation concepts needs to handle that limitation. Even if a complete toolbox would ever be achievable, until we get there, we need a back-up.
  
\subsubsection{Narrow Scope}
As Ceccato et al.\ observed~\cite{emse2019}, there exist multipe options to overcome SPs, deobfuscation being only one of them, and likely the hardest. While this does not diminish the relevance of the existing studies of automated deobfuscation, it does put into question the focus of resilience on deobfuscation, and the corresponding neglecting of alternatives to defeat SPs by bypassing them, overcoming them, or building workarounds. Ceccato et al.'s taxonomy describes these alternative strategies.

\subsubsection{Attack Oversimplification}
Another way attacks are often narrowed down to deobfuscation, and therefore oversimplified, is in their neglect of how to locate the code to be deobfuscated. One class of papers that suffers from this is papers focusing on deobfuscation of mixed Boolean-arithmetic (MBA) expressions. The vast majority of those papers, such as~\cite{2021_mba_blast_unveiling_and_simplifying_mixed_boolean_arithmetic_obfuscation}, only study the deobfuscation of symbolic MBA expressions but sidestep the problem of how to identify deobfuscatable expressions in binaries.

De Sutter et al.~\cite{desutter2024evaluation} also observed that most obfuscation publications only consider one individual analysis method in their tool-based evaluation, if they have one at all. Clearly, guidance is needed towards considering real-world attacks in which multiple methods are combined. Recently, tools and models for doing so have been proposed~\cite{checkmate24}, which may be a first step in this direction. 

\subsubsection{Unvalidated Metrics}
\label{unvalidated_metrics}
Just as many software complexity metrics have not been validated on obfuscated code (see Section~\ref{sec:unfit_complexity_metrics}), none of the ad hoc tool-based metrics have been validated, i.e., the relation between those metrics and actual attack effort or success rate has not been determined. For example, Van den Broeck et al.~\cite{jens21} count how many edges are drawn by IDA Pro in functions' control flow graphs, but is it really relevant whether or not an edge that IDA Pro knows about and records in its database of CFGs, is drawn on screen or not? This has not been validated at all. 

This criticism is harsh, as it is probably not feasible to validate all relevant metrics with empirical experiments because those are too expensive. Authors reporting tool-based metrics should hence probably be forgiven. Still, it is a threat to validity. 

\subsubsection{Buggy and Incomplete Tools}
\label{sec:buggy_tools}
Besides the impact of tool availability (see Section~\ref{sec:tool_availability}) and of their limitations and peculiar heuristics (see Sections~\ref{sec:unclear_computations} and~\ref{sec:adhoc}), an additional issue is that software analysis tools invariably have bugs. Commercial protection tools sometimes exploit these, e.g., by injecting instructions into a binary that cause certain tools to crash. Academic analysis tools are obviously also often buggy in the sense that their implementation will typically be incomplete. As research tools, their authors might very well have cut some corners to make the tools work on their use cases and benchmarks, but not beyond them. 

Obviously, SP strength evaluations should be transparent with respect to results depending on bugs or completeness issues with the used tools. Although we know of no papers published in top conferences or journals that violate this requirement overtly, we have reviewed (and blocked from publications) papers making this mistake multiple times in the past. So we think this issue deserves explicit handling in the guidelines. 

\subsubsection{Confusion about Potency and Resilience}
The use of the term resilience by Cozza et al.\ does not correspond to CTL's original definition, as it does not target automated deobfuscation. This is in line with what was already mentioned in Section~\ref{sec:fuzzy_boundary}, namely that researchers sometimes claim to evaluate, assess, or predict resilience because they study automated techniques, but in fact do not study deobfuscation techniques~\cite{banescu15,2017_predicting_the_resilience_of_obfuscated_code_against_symbolic_execution_attacks_via_machine_learning}, so whether they really study potency or resilience is up for discussion. In this case, the question is whether computing invariants can be seen as deobfuscation. Clearly the existing definitions are confusing. 

To some extent, this seems unavoidable, because some analysis can be used as part of deobfuscation, but also exist outside that context. Another example is the computation of points-to set sizes, as done by Foket et al.~\cite{Foket14}. Computing points-to sets (a.k.a.\ as class analysis~\cite{vortex}) is necessary to compute precise call graphs of software written in object-oriented programming languages that feature polymorphism. 
It is hence an analysis that clearly exists outside of deobfuscation. But when an obfuscation aims precisely at making call graphs imprecise, points-to set computation is part of the deobfuscation. Evaluating the impact of that obfuscation on points-to sets hence seems to touch both potency and resilience.

The use of plug-ins by Van den Broeck et al.~\cite{jens21} provides another example. Those plug-ins try to fix some of the worst performing heuristics of IDA Pro's CFG reconstruction approach when facing their novel obfuscation. Is an evaluation with such plug-ins to be seen as evaluating that obfuscation's potency or its resilience? Again, one evaluation seems to touch on both those criteria. Trying to draw a strict line between them is hence perhaps futile. 

\section{A New Framework}
The formulated critiques articulate that a better formulation of evaluation criteria for obfuscating transformations is needed to provide better guidance on how to assess their real strengths and weaknesses rather than reporting artificial, irrelevant, or even misguided proxies. The primary aim of such guidance is to help researchers maximize the impact of their work by convincing other researchers that they can build on this work for their own research and by convincing practitioners of the relevance of the work for real-world software protection. To a large degree, such impact depends on the validity of the performed evaluations. This includes the commonly used criteria of validity in software engineering, such as construct, conclusion, internal, and external validity~\cite{Wohlin} but also less commonly considered criteria such as instantiation validity~\cite{lukyanenko2014instantiation}. \emph{Reducing such threats to validity is the aim of this framework.} 

In line with many of the existing definitions discussed in Sections~\ref{sec:collberg_nagra_def},~\ref{sec:completeness}, and~\ref{sec:adequacy}; in line much of the examples mentioned in Section~\ref{sec:attack_tool_examples}; and in line with the recommendations of De Sutter et al.~\cite{desutter2024evaluation} that evaluations obfuscation strength evaluations should be based on their impact on real-world attacks, we put forward that the used evaluation criteria should be based on specific program properties that adversaries can target with specific attack strategies, and on the impact that the obfuscations have on those strategies. 

Unlike NC, we think that a concept similar to resilience needs to be a core criterion. And unlike the research on AI-based potency, we feel the criteria need to be practical, e.g., in the sense that they are consistent with the fact that MATE attackers can always reveal the properties they are after, and that the required effort of a whole attack strategy, not only of individual steps, is what matters. Unlike existing evaluation criteria defined in terms of analysis, we define them in terms of attack strategies, which consist of different categories of attacks steps being executed: 
\begin{enumerate}
\item knowledge gathering, i.e., revealing properties:
\begin{enumerate}
    \item revealing static/dynamic artifacts in the program;
    \item revealing relations between artifacts;
    \item revealing features of the artifacts and the relations;
    \item assigning priorities to artifacts and relations; 
    \item revealing mappings between abstract artifacts, properties, and relations; and concrete ones.
\end{enumerate}
\item artifact manipulation, such as lifting a code fragment from a program for ex situ execution, tampering with a SP to undo it, or altering its execution state to bypass a SP; 
\item decision making on the next steps to execute based on the determined priorities and already gathered knowledge.
\end{enumerate}
These categories capture the four goals of analyses in the survey by Schrittwieser et al.\ on obfuscation vs.\ program analysis~\cite{survey2016}, the activities modeled by the reverse engineering formalization by Faingnaert et al.~\cite{checkmate24}, the activities observed in experiments by Ceccato et al.~\cite{emse2019}, in a survey on the practice of malware analysis~\cite{wong2021inside}, etc. For example, code comprehension tasks, such as understanding that some code fragment implements a quicksort, are instantiated by combinations of (1a), (1b), (1c), and (1e). The same holds for making hypotheses, confirming them, and discarding them, which are important attack activities as observed by Ceccato et al.~\cite{emse2019}. Finding the location of relevant code and data in a program in addition requires (1d) to classify the revealed information into relevant vs.\ irrelevant ones. Undoing an obfuscation obviously requires (2). 

\emph{It is hence against combinations of these kinds of attack steps, and the possible methods to execute them in relevant attack strategies, that obfuscations can and should be evaluated.}

Importantly, the methods used to execute these steps can be unsound and non-conservative. For example, trace-based analysis techniques to reveal properties are unsound. Artifact manipulation may be non-conservative when the applied transformation does not conserve the behavior under all possible circumstances. When some hypothesis is being made by an adversary in some attack steps that are not confirmed in later steps, for example because the adversary reaches their goal first, the steps building on that hypothesis are obviously unsound as well. 

Given the wide range of properties that adversaries might be interested in, and the wide range of analyses techniques that can help them, it makes no sense to propose a one-size-fits-all set of evaluation metrics or to prescribe exactly what metrics should be measured. We hence limit ourselves to prescribing which criteria should be evaluated, aiming at defining them in ways that avoid the confusion that has existed regarding previous definitions. Our new framework of criteria includes relevance, effectiveness (and its poor man's alternative of efficacy), robustness, concealment, stubbornness, sensitivity, predictability, and cost. Table~\ref{tab:criteria} lists them all, including their subcriteria. 

\begin{table}[t]
    \centering
    \begin{tabular}{c p{6cm} l}
\hline
    \multicolumn{3}{l}{\textbf{Relevance}}\\
       & Attack Steps Relevance  & $Re_{a}$ \\
       & Program Property Relevance & $Re_{p}$ \\
       & Obfuscation Impact Relevance & $Re_{o}$ \\
       & Metrics Relevance & $Re_{m}$ \\
       & Tool Availability:& $Re_{t}$ \\
       & Sample Relevance & $Re_{s}$\\
       & Layered Protection Relevance & $Re_{l}$\\
    \multicolumn{3}{l}{\textbf{Effectiveness}}\\
   & Isolated Outcome Effectivness & $E_{o,i}$ \\
   & Marginal Outcome Effectivness & $E_{o,m}$ \\   
   & Isolated Resource Effectivness & $E_{r,i}$ \\
   & Marginal Resource Effectivness & $E_{r,m}$ \\
    \multicolumn{3}{l}{\textbf{Robustness}}\\
     &  Isolated Outcome Delta & $Ro_{o,i}$ \\
     &  Marginal Outcome Delta & $Ro_{o,m}$ \\
     &  Isolated Resource Delta & $Ro_{r,i}$ \\
     &  Marginal Resource Delta & $Ro_{r,m}$ \\
     &  Deployment Delta & $Rd_{o,m}$ \\
    \multicolumn{3}{l}{\textbf{Concealment}}\\
         &  Local Concealment & $C_{l}$ \\
         &  Global Concealment & $C_{g}$ \\
         &  Strategic Concealment & $C_{s}$ \\
    \multicolumn{3}{l}{\textbf{Stubbornness}}\\
         &  Outcome Stubbornness & $St_{o}$ \\
         &  Resource Stubbornness & $St_{d}$ \\
   \multicolumn{3}{l}{\textbf{Sensitivity}}\\
         &  Sample Feature Sensitivity & $Se_{s}$ \\
         &  Attack Instantiation Sensitivity & $Se_{a}$ \\       
         &  Protection Instantiation Sensitivity & $Se_{pi}$ \\       
         &  Protection Configuration Sensitivity & $Se_{pc}$ \\       
         &  Build Tool Flow Sensitivity & $Se_{b}$ \\                
         &  Platform Sensitivity & $Se_{p}$ \\ 
   \multicolumn{3}{l}{\textbf{Predictability}}\\
         &  Sample Feature Predictability & $P_{s}$ \\
         &  Attack Instantiation Predictability & $P_{a}$ \\       
         &  Protection Instantiation Predictability & $P_{pi}$ \\       
         &  Protection Configuration Predictability & $P_{pc}$ \\       
         &  Build Tool Flow Predictability & $P_{b}$ \\                
         &  Platform Predictability & $P_{p}$ \\ 
   \multicolumn{3}{l}{\textbf{Cost}}\\
         &  Performance Cost & $C_{p}$ \\
         &  SDLC Cost & $C_{sdlc}$ \\       
         \hline
    \end{tabular}
    \caption{All obfuscation evaluation criteria and subcriteria}
    \label{tab:criteria}
\end{table}

\subsection{Relevance of the Evaluation Constructs}
The first criterion to be considered in an evaluation of an obfuscation's strength is the relevance of all the constructs used in the evaluation. To maximize the relevance of a researcher's evaluation results for practitioners, the used constructs need to representative for real-world software protection.

Researchers, however, face deadlines and budget and resource limitations. Different researchers therefore aim for different technology readiness levels (TRLs) with their research. That is obviously fine, and different criteria of our framework support research and evaluations at the various levels, while at the same time enabling and encouraging the researchers to be transparent about the TRL of their research results. Ensuring this transparency, and forcing the researcher to reason about the impact of evaluation methodology choices on the relevance of the evaluation results and their validity is the aim of the seven relevance criteria detailed below. 

It should come as no surprise that we first focus on criteria that pertain more to the evaluation method than to the obfuscation being evaluated itself. The reason is of course the lack of standardization in the domain of SP~\cite{Basile23}. Where certification has been standardized (e.g., smart card security~\cite{common_criteria}) and where risk management standards have been adopted (e.g., NIST SP 800-39~\cite{nistSP800-39} in network security~\cite{Gartner-report-riskanalysis}), concise evaluation results can be published that refer to those standards so that all stakeholders know how to interpret those results. In MATE SP, on the contrary, only some embryonic steps toward standardization have been proposed~\cite{Basile23}. Lacking standardized evaluation methods, evaluation results are meaningless without proper framing. 

An evaluation therefore stands or falls with accurate and complete framing, which can be achieved by assessing the evaluation according to the seven relevance criteria we put forward:

\paragraph{$Re_a$ - Attack Steps Relevance:} To what extent are the considered attack step combinations relevant? In which relevant attack strategies are they used, meaning attack strategies with a proven track record in the scientific literature or in real-world reporting? Are there no alternative combinations known that can produce the same results as the considered ones but are simpler to execute? Is the starting point of the considered attack steps realistic, i.e., the preceding attack steps? Does the evaluation use realistic outputs of those preceding steps as inputs to evaluate the considered attack step combination?

\paragraph{$Re_p$ - Program Property Relevance:} To what extent are the program properties relevant that the protection is supposed to hide, and that adversaries supposedly want to reveal through the considered attack steps? Are there alternative strategies with which adversaries can reach the same end goal, but without requiring them to obtain the results of the considered attack steps, i.e., without them requiring to reveal exactly those program properties?

\paragraph{$Re_o$ - Obfuscation Impact Relevance:} To what extent does the impact of the evaluated obfuscation on the result of the considered attack step combination affect the execution of later attack steps? Does it make them impossible? Does it make them require more resources? Does it make them produce less precise results? 

\paragraph{$Re_m$ - Metrics Relevance:} Do the (commonly used or ad hoc) metrics used to assess the obfuscation's impact on the considered attack steps' outputs truly capture the impact on later attack steps? Has that been validated, or are there strong arguments or evidence? 

\paragraph{$Re_t$ - Tool Availability:} Is practical tool support available to automate the considered attack steps and strategy? 

\paragraph{$Re_s$ - Sample Relevance:} Are the used samples representative for the types of software that would be attacked with the considered attack steps and strategies in the real-world?  

\paragraph{$Re_l$ - Layered Protection Relevance:} Does the evaluation consider relevant layerings or compositions of SPs?  

\vspace{0.2cm}

Note that some criteria are related, such as $Re_a$ and $Re_t$. It is the task of the researcher to be consistent in their assessment of their evaluation with respect to these criteria.  Also note that these criteria are to be evaluated mostly, if not completely, qualitatively. 

\subsection{Effectiveness against Standard Attacks}
\label{sec:effectiveness}

The effectiveness of an obfuscation for protecting a secret property in a given program against a considered attack step combination is defined as the obfuscation's effect on that combination, for that program and that property. 

To a large degree, effectiveness equals the old criterion of potency as defined by NC (see Section~\ref{sec:collberg_nagra_def}). There is one major difference, however: We specifically prescribe that the criterion effectiveness should only be used when the considered attack steps are \emph{standard attack steps}, i.e., interesting options for an adversary with no a priori knowledge about the specific SPs being deployed to counter the considered attack steps. In other words, these are the attack steps that can be expected to work well on vanilla applications as well as on the average protected application that the adversary may expect to be facing.  It is because of this crucial difference that we propose not to reuse and redefine the term potency, but instead put forward the term effectiveness. 

The notion of a standard attack steps is deliberately fuzzy. It is up to each researcher to determine which attacks steps to consider standard. The use of the term effectiveness then comes with the responsibility  to argue why the considered steps are to be considered standard. This is of course closely related to the relevance criteria $Re_a$, $Re_t$, and $Re_o$. It is the researcher's responsibility to ensure consistency in their assessment. 

Just like NC's potency considers the impact of an obfuscation in terms of analysis outcomes, and the required resources, so does effectiveness entail two criteria:

\paragraph{$E_o$ - Outcome Effectiveness} What is the effect of the obfuscation on the outcome of the standard attack step combination? Does the result become more complex, less precise, less accurate, incorrect, etc.? Often this effect can be quantified with (ad hoc) metrics. 

\paragraph{$E_r$ - Resource Effectiveness} What is the effect on the required resources to execute the standard attack steps. This effect can often be quantified. Many different forms of resources can be considered: computation time, memory footprint, network bandwidth, etc.

\vspace{0.2cm}

The effectiveness of an obfuscation for protecting a class of secret properties against a considered attack step combination is then to be obtained by evaluating it on multiple samples and by reporting the distribution of the obtained results. 

The metrics used to assess the outcome effectiveness $E_o$ should be computed on results produced by actual attack steps or good proxies thereof, not on ground-truth data produced while building the protected samples. So they should be tool-based, like the examples in Section~\ref{sec:attack_tool_examples}. Complexity metrics originating from the domain of software engineering can still be used, on program representations reconstructed by attack tools such as disassemblers, if there exists a correlation between the used metrics and the execution of later attack steps in the considered attack strategies. Obviously this relates to the relevance criterion $Re_m$. 

Of both the criteria, two forms can be considered. Isolated effectiveness ($E_{o,i}$ and $E_{r,i}$) is the effect of an obfuscation when it is evaluated in isolation. Marginal effectiveness ($E_{o,m}$ and $E_{r,m}$) is the effect of deploying the obfuscation in combination with other (commonly used) SPs. It is strongly encouraged to evaluate marginal effectiveness. For example, when evaluating the impact of some obfuscation on Javascript code, it should be evaluated on top of basic minification transformations~\cite{2019_anything_to_hide_studying_minified_and_obfuscated_code_in_the_web} that any Javascript obfuscator supports, such as identifier renaming~\cite{liu2017stochastic}. 

Finally, we stress that if an obfuscation aims at complicating multiple, different attack steps or step combinations, its effectiveness should also be evaluated for each of those steps or combinations, resulting in multiple results for outcome and resource effectiveness. 

\subsection{Efficacy - a Poor Man's Effectiveness}
\label{sec:efficacy}
When a researcher cannot provide convincing arguments or evidence for the considered attacks steps being standard and relevant, or when no actual attack steps are used in an evaluation, such as when only complexity metrics are being computed on ground-truth data, the term effectiveness should not be used. At best, the term \emph{efficacy} (with shorthands $e_o$ and $e_r$ instead of $E_o$ and $E_r$) can then be used to stress that something akin to lab conditions is being evaluated, rather an evaluation indicative for real-world conditions. 

\subsection{Robustness against Special-purpose Attacks}
As effectiveness considers only standard attacks steps, we need a counterpart for non-standard attack steps. For this purpose, the robustness of an obfuscation is determined in terms of how much better \emph{special-purpose analyses} can do than standard analyses. Better can be in terms of required resources and in terms of produced result. Special-purpose attack steps are those that perform (on average) worse than standard ones on samples that do not feature the evaluated obfuscation, and that would therefore not be chosen by adversaries unless they know they are attacking this obfuscation. 

Special-purpose attack steps benefit attackers if they require less resources or produce better results than standard ones on at least some programs protected with the obfuscation being evaluated. An example is the $k$-depth interprocedural constant propagation~\cite{interp_cp} discussed in Section~\ref{sec:perspective}, where $k$ is increased only for those parts of the program where it matters to reveal the secret property. Another example is a devirtualization technique tweaked for specific forms of virtualization~\cite{deobf_virtualization,kinder}.

The robustness against some special-purpose attack step combination is evaluated in three dimensions:

\paragraph{$Ro_o$ - Outcome Delta} How much worse or better is the result of some special-purpose attack on a protected sample compared to the result of a standard attack on it?

\paragraph{$Ro_r$ - Resource Delta} How much less or more resources does the special-purpose attack require on that sample?

\paragraph{$Ro_d$ - Deployment Delta} How much more configuration, customization, or development effort does the adversary need to invest before to enable the special-purpose attack, for example because no out-of-the-box tool support is available?

\vspace{0.2cm}

Obviously $Ro_o$ and $Ro_r$ should ideally be reported as distributions obtained on a number of samples.

Just like we did for effectiveness, we put forward isolated and marginal forms of these robustness criteria, to distinguish between the robustness of an obfuscation deployed in isolation, and the gain in robustness when deploying an obfuscation on top of other SPs. 

The deployment delta $Ro_d$ can be considered the updated equivalent of the developer effort in the definition of resilience by CTL (see Section~\ref{sec:resilience}). Unlike them, we do not prescribe the specific features that should be taken into account, nor do we prescribe a specific scale of robustness levels. We leave it up to the researchers to choose an appropriate assessment method. 
Obviously, $Ro_d$ and $Re_t$ are closely related. It is up to the researcher performing the evaluation to ensure consistency.

\subsection{Concealment}
Concealment is the new framework's equivalent of the older concept of stealth. As discussed in Section~\ref{sec:narrow_stealth}, not only the accuracy of detector and locator functions needs to be assessed. A complete assessment of an obfuscation needs to include all its aspects that can help an adversary make strategic decisions. Concealment hence needs to be, and is, a more generic concept than the old stealth. 
 
One of the most important strategic decisions relates to robustness. The relevant question in that case is the following: If an obfuscation lacks in terms of stubbornness because special-purpose attack steps can defeat it, how easy is it for an adversary to determine which special-purpose attacks to choose? In other words, how likely will the adversary be able to decide to use that special-purpose attack. As an example, remember the discussion in Section~\ref{sec:perspective} on how adversaries can or cannot choose refinements of attack steps, such as choosing on which parts of the program to use a higher value of $k$ for $k$-depth interprocedural constant propagation. If an adversary has no way of determining these parts, i.e., if the adversary cannot create an oracle to produce the required configuration for such a special-purpose analysis, the defender should probably not be worried about its existence. 

To capture all relevant forms of information about the deployed protections, we put forward three concealment criteria: 

\paragraph{$C_l$ - Local Concealment} To what extent can a  locator function determine the exact locations in a program where an obfuscation has been deployed. This is the equivalent of NC's local stealth. We strongly encourage the use of standard accuracy metrics, however, instead of their proposal to compute the maximum of the false positive and false negative rates. 

\paragraph{$C_g$ - Global Concealment} To what extent can a detector function determine whether an obfuscation has been applied on a program. This is equivalent to NC's steganographic stealth. Again we encourage using standard accuracy metrics instead of a max function. 

\paragraph{$C_s$ - Strategic Concealment} To what extent can an adversary reveal additional information, of any kind, about the deployed protection that can drive strategic decisions in the attack strategy, in particular to determine which potentially beneficial special-purpose attack steps to execute next. We foresee that this criterion will most often be assessed qualitatively.

\vspace{0.2cm}

Note that as we did for effectiveness to replace potency, we opted to use the term concealment to replace stealth, even though they are very similar. We made this choice to avoid confusion when researchers report strength evaluations in the future. 

\subsection{Stubbornness against Deobfuscation}
Undoing an obfuscation is only one way to defeat it~\cite{emse2019}. It is still an important one, however, so the ease with which an obfuscation can be undone should be considered when evaluating it. Here the term ``undoing'' refers specifically to applying a transformation that undoes the effect of an obfuscation. It excludes performing the necessary analyses to determine which transformation to apply. 

Consider, for example, an opaque predicate that is always true and that is used in a conditional branch, which is hence always taken. With some preceding attack step ---of which the effectiveness can of course be evaluated--- the attacker might have determined that this predicate is always true. Stubbornness does not concern that preceding step, it only concerns the ease with which the opaque predicate insertion can be undone. With the code patching functionality of interactive disassemblers and decompilers such as Ghidra, this is trivial: it suffices to replace the conditional branch by a conditional one, thus simply removing the bogus execution path. Ghidra then automatically eliminates the now dead predicate computation and the bogus control flow path and bogus code from the reconstructed CFG and from the decompiled code. So a simple code patch suffices to completely undo the obfuscation. 

For other obfuscations, that are in nature perhaps not more advanced than opaque predicates, such easy undoing edits are not available. For example, after control flow flattening each flattened block has only one successor in the CFG, namely the dispatcher of the flattened code~\cite{flattening}. For reverting to the original CFG, simple edits that remove an execution path from the code will not suffice. 

To capture the ease with which such obfuscation-undoing edits and transformations can be performed on the representation of the software on which later attack steps will operate, we define two stubbornness criteria: 

\paragraph{$St_o$ - Outcome Stubbornness} To what extent can the effect of the obfuscation be undone in a relevant representation of the code? 

\paragraph{$St_d$ - Resource Stubbornness} How difficult is it to perform that deobfuscating transformation? 

\vspace{0.2cm}

For both aspects, only a qualitative assessment can be expected, which will probably involve ad hoc arguments and evidence. We certainly cannot prescribe specific methods at this point in time. But obviously, this assessment will need to be consistent with how other criteria are assessed, such as $Re_t$.

Note that multiple representations might need to be considered for these stubbornness criteria. If the later attack steps are assumed to be static code comprehension, editing Ghidra's internal representation of the binary code can suffice, as explained above for opaque predicates. If the follow-up attack steps to the contrary involve dynamic analysis techniques for which a patched binary needs to be executed, changing a tool's internal representation of a program will not suffice. In that case, a working edited binary actually needs to be produced first. Ghidra can generate a patched binary from its internal representation, but whether that binary will execute correctly will depend, among others, on the presence of anti-tampering protections such as remote attestation~\cite{viticchie2016reactive}. As was the case for effectiveness, also for stubbornness isolated and marginal stubbornness criteria can hence be considered.

\subsection{Sensitivity}
\label{sec:sensitivity}
Section~\ref{sec:effectiveness} explicitly broadened the criterion of effectiveness from the impact observed one program sample regarding one property to effectiveness on a class of properties on many samples. For the other criteria, a similar broadening can, of course, be done as well. If an obfuscation's strength for any criterion is sensitive to certain features of the samples being protected, and the samples exhibit sufficient variance with respect to those features, this sensitivity will show up in the resulting distribution reported for that criterion. 

Regardless of whether or not such a distribution already provides evidence of such sensitivity, we encourage researchers to reflect on it explicitly. They can do so by providing evidence that the used samples cover a sufficient spread of the relevant features, or by arguing qualitatively how sensitive the evaluated obfuscations and the evaluation criteria are to certain program features. 

Such a reflection should not be limited to the features of the programs to be protected however. To the contrary, sensitivities to several confounding factors should be assessed:

\paragraph{$Se_s$ - Sample Feature Sensitivity} To what extent are the other criteria sensitive to features of the programs to be protected?

\paragraph{$Se_{a}$ - Attack Instantiation Sensitivity} To what extent are the evaluation results for other criteria sensitive to design and implementation details of the considered attack steps? An example of such details are the precise data flow analyses that are used to implement certain attack steps, or the specific tools that are used thereto. This criterion is particularly important when the evaluation relies on research prototypes of software analysis techniques. As we stated in Section~\ref{sec:buggy_tools}, it is not common for such tools to be incomplete or buggy. If the evaluation results depend on bugs or missing features, this clearly needs to be reported. 

\paragraph{$Se_{pi}$ - Protection Instantiation Sensitivity} To what extent are the evaluation results for other criteria sensitive to the implementation and design details of the (research) tools developed to deploy the obfuscation on samples? Obviously, it are not only analysis tools developed by researchers that can be incomplete, so can research SP prototypes; hence the inclusion of this criterion. This criterion is particularly relevant for research into deobfuscation and software analysis methods (including malware detection), in which researchers have to evaluate to what extent their novel techniques can defeat variations of obfuscations rather than being overfitted to particular variants.  

\paragraph{$Se_{pc}$ - Protection Configuration Sensitivity} To what extent are the evaluation results for other criteria sensitive to the configuration parameters of the deployed protections? Tigress~\cite{tigress2023}, for example, offers a wide range of configuration options for each supported obfuscation. It should at least be clear which configurations have been used in the evaluation. Of particular interest are random seeds used to drive stochastic SP techniques that many SP tools support. Such techniques are often used to make obfuscations renewable, meaning that they do look somewhat different every time they are deployed. If the evaluation results are sensitive to the used parameters or random seeds, i.e., if they display a large variability, the researchers ideally perform a parameter sweep and measure the resulting variance in the evaluation criteria. 

\paragraph{$Se_{b}$ - Build Tool Flow Sensitivity} To what extent are the evaluation results for other criteria sensitive to the build tools with which the protected application is being built? The interaction between SP transformations and other transformations applied during the build process of a protected application, such as compiler optimizations that are executed after source-to-source rewriting has been applied to inject obfuscations (e.g., with Tigress~\cite{tigress2023}) or after compile-time protection passes have been executed (e.g., with OLLVM~\cite{ollvm} or Epona~\cite{epona}) are complex~\cite{optimization}. Obfuscation transformations risk to be undone completely or partially by clever compiler optimizations, as discussed in Section~\ref{sec:sensitivity_missing}. The sensitivity of the evaluation results to such interactions hence needs to be assessed. 

\paragraph{$Se_{p}$ - Platform Sensitivity} To what extent are the results limited to or dependent on the platform for which the software is built, such as the operating system, the processor architecture, etc.

\vspace{0.2cm}

Most of these criteria can in theory be assessed by performing sweeps over sufficiently diverse samples, and sufficient configurations and implementation choices, of the attack steps and of the obfuscations. In practice, however, researchers most likely will not have the time and resources to experiment with all potentially relevant configurations and implementations. We therefore expect several aspects to be assessed qualitatively, and in ad hoc manners. 

Beyond helping to reduce the threats to validity, these sensitivity criteria are critical for the user-friendliness of the tools supporting the obfuscation. As argued in Section~\ref{sec:sensitivity_missing}, the lower the different forms of sensitivity of an obfuscation, the more predictable its strength will be for different programs, for future evolutions on the same program, and/or in light of (future) evolutions of program analysis techniques and attacker capabilities. 

Note that $Se_s$ and $Se_{pi}$ combined cover what can be called the applicability of an obfuscation. Many obfuscations can only be applied to certain types of software artifacts, or when certain preconditions are met. Some limitations of an obfuscation may be fundamental, in other cases a prototype instantiation has limited applicability because of a lack of resources to engineer a more complete implementation. That is fine, but the difference between fundamental limitations and instantiation should be made clear in any evaluation. The criteria $Se_s$ and $Se_{pi}$ serve that purpose.  

\subsection{Predictability}
The fundamental reason why low sensitivity to program features can be beneficial as argued above, is that it makes the result of an obfuscation more predictable. Higher predictability can benefit a SP optimization process for selecting the best combination of SPs given a program, its assets, and their security requirements. Similarly, if the impact on variations of attack steps can easily be predicted, it can become easier for decision support tools~\cite{checkmate24} and for predictive ML models~\cite{reganoMetric,2017_predicting_the_resilience_of_obfuscated_code_against_symbolic_execution_attacks_via_machine_learning} to predict the impact of obfuscations on a range of similar attack steps.  

For these reasons, we put forward complementary predictability criteria $P_s$, $P_{a}$, $P_{pi}$, $P_{pc}$, $P_{b}$, and $P_{p}$ that have similar definitions as their sensitivity counterparts, but for predictability instead of sensitivity. Similar to the sensitivity criteria, we expect these to be assessed mostly, if not completely, qualitatively. 
 
\subsection{Cost}
A detailed discussion of the various performance criteria that can be useful for evaluating an obfuscation is out of scope. They can include network bandwidth, client-side execution time and memory footprint, server-side execution time and memory footprint, power consumption, real-time behavior, response latency, throughput, and various other performance criteria. Which ones are relevant depends on the program to protect and its non-security-related non-functional requirements. We hence do not prescribe any specific criterion, but we do observe that in literature~\cite{desutter2024evaluation}, static program size and execution time are by far the most evaluated criteria, followed at a great distance by compilation time, dynamic memory footprint, and power consumption. 

We do stress, however, that when a specific performance criterion is relevant, it is important to evaluate its sensitivity as discussed in Section~\ref{sec:sensitivity}. In particular, the sensitivity to the hotness of protected program fragments should be assessed, i.e., to their execution frequency and relative contribution to the overall performance.

Moreover, the cost of using certain obfuscations can extend beyond the performance criteria. Various (expensive) processes in the whole software development life cycle (SDLC) can be impacted by the use of obfuscations, such as certification, quality assurance, debugging, distribution, etc. 

To capture both forms of costs, we put forward two broadly defined cost criteria:

\paragraph{$C_{p}$ - Performance Cost} How is the quantitative overhead of the protections in terms of all kinds of performance costs distributed? 

\paragraph{$C_{sdlc}$ - SDLC Cost} What are, qualitatively or quantitatively, the impacts that the deployment of the evaluation obfuscation can have on the SDLC? Where can the deployment of the obfuscation pose challenges or compatibility issues with industrial SDLCs?

\section{Discussion}
\label{sec:discussion}

The proposed framework comprises more than 30 evaluation criteria in eight categories, plus one backup category. This framework and its criteria aim for complete coverage of the relevant criteria, rather than for prescribing evaluation methods in detail. It aims for informal guidelines that help practitioners limit the threats to the validity of their research, rather than for formal definitions. 

Individual researchers that will (hopefully) use this evaluation framework still have considerable degrees of freedom to instantiate the proposed criteria and a responsibility to do so adequately. By doing so, they will be able to maximally mitigate threats to internal validity~\cite{Wohlin} and conclusion validity~\cite{Wohlin}. 

By emphasizing relevance, layered use of protections, evaluations of the impact on adversaries' attack steps, and various forms of sensitivity, the proposed framework can help researchers mitigate threats to external validity~\cite{Wohlin}, to construct validity~\cite{Wohlin}, and to instantiation validity~\cite{lukyanenko2014instantiation}.

Of particular interest is that our framework allows, and even encourages to use ad hoc tool-based metrics, which at first sight might seem to contradict the critique formulated in Section~\ref{sec:adhoc} regarding the use of ad hoc metrics and ad hoc tool configuration. Given the wide range of program properties that adversaries might be interested in, the wide range of program features that can impact the adversaries' effectiveness and efficiency in executing their attack strategies, the wide range of analyses techniques and tools that can help them, as well as the various alternative algorithms and heuristics such tools can build on, in combination with the need focus on impacting actual attack steps instead of artificial metrics, we see the use of ad hoc tool-based metrics as unavoidable. In fact, if done well, we see it as absolutely beneficial. 

Multiple criteria focus on using ad hoc tool-based metrics adequately. Combined, the criteria attack steps relevance, program property relevance, and metrics relevance aim at ensuring that measured ad hoc features and metrics are relevant. Similarly, the criteria attack steps relevance, robustness, and the attack instantiation sensitivity encourage researchers to consider how adversaries could adapt their existing attack strategies and heuristics, i.e., to consider not only the status quo but also the next steps in the cat and mouse game between defenders and attackers.

Over time, we hope that the use of this framework itself will become a best practice, but also that a catalogue of best practices will be built as new and better instantiations of them get published. 

Of all the critiques formulated in Sections 2--5, the only ones not satisfactorily addressed by our framework are those on software complexity metrics from the domain of software engineering not necessarily being valid when used on obfuscated code (Sections~\ref{sec:unfit_complexity_metrics} and~\ref{unvalidated_metrics}). We can only recommend that more empirical research is performed to validate metrics for the assessment of reverse engineering activities on obfuscated code, such as~
\cite{emse2019,2014afamily} to improve the relevance of such complexity metrics. Notice that even in the domain of software engineering, they are contested, and more empirical research on their validation is required~\cite{feitelson}. In our framework, we discourage their use by prioritizing the use of effectiveness criteria (Section~\ref{sec:effectiveness}) over efficacy criteria (Section~\ref{sec:efficacy}). 

\section{Related Work}

This work follows directly in the footsteps of the 2019 Dagstuhl Seminar on SP Decision Support and Evaluation Methodologies~\cite{Dagstuhl} and the 2024 survey on Evaluation Methodologies in Software Protection Research~\cite{desutter2024evaluation}.

\subsection{Dagstuhl Seminar 19331}
The report on Dagstuhl Seminar 19331~\cite{Dagstuhl} includes two case studies, for which recommendations were formulated following a number of brainstorm sessions with all seminar participants. The first use case is anti-disassembly protection. The second use case is trace-based analysis, which obviously can play an important role in attacks on obfuscated code. For both use cases, the report discusses viable and relevant attack scenarios as well as relevant sensitivities to program features and other factors that researchers should consider. We can confirm that the evaluation criteria proposed in this paper capture all recommendations that have been formulated for those use cases. 

That does not imply, however, that with this work we cross the finish line of the search into better evaluation methodologies that started at Dagstuhl. Most importantly, the SP research domain is still in search of standardized benchmark suites for use in software protection research. Building such suites of unprotected and protected samples is orthogonal to the ambitions and scope of this paper. Another open challenge is to develop an experimental environment that as is as complete as possible in terms of the attack strategies of which it automates the execution/simulation/assessment, and in which such samples can easily be evaluated in a reproducible way. This aspect was already discussed in Section~\ref{sec:adhoc}. Some first attempts have been recently made in that direction~\cite{checkmate24}, but that work is also complementary to this paper, as it focuses on how to develop models and effort estimation methods for reverse engineering attacks, whereas this paper focuses on how to properly use such models and methods.  

\subsection{Survey on Evaluation Methodologies}

This 2024 survey formulated ten recommendations~\cite{desutter2024evaluation}, which are all reflected in one way or another in the criteria proposed here.

\paragraph{Multiperspectivism} De Sutter et al.\ first recommend that when an evaluation of a new obfuscation should included an analysis of basic adaptations that an attacker might make to their existing analyses as a countermeasure. The explicit focus on standard analyses vs.\ special-purpose analyses in the definitions of effectiveness and robustness respectively aim, in part, at capturing this. Whether adaptions by adversaries should be presented as standard or special-purpose attack steps is left open: it depends on whether the adaptations are generally applicable or whether their use should be limited to scenarios in which the evaluated obfuscation has already been detected by the adversary. 
\paragraph{Complete strength evaluation} De Sutter et al.\ then recommend evaluating the impact of obfuscations on attacks on obfuscated assets themselves as well as attacks on the obfuscation, including its stealth. While they still reuse the original terminology of potency, resilience, and stealth, their recommendation is of complete strength evaluation is precisely what this framework is about. The criteria relevance, effectiveness, robustness, concealment, and stubbornness more than cover their second recommendation.  
\paragraph{Layered SP deployment} De Sutter et al.\ recommend exactly what we recommend in Section~\ref{sec:effectiveness} on evaluation the marginal outcome and resource effectiveness $E_{o,m}$ and $E_{r,m}$. 
\paragraph{Concrete attacks evaluation} Our focus on attack step combinations, relevance, and the use of tools to evaluate effectiveness, robustness, and stubbornness instantiate the recommendation of focusing on concrete attacks rather than artificial metrics. 
\paragraph{Sample diversification} De Sutter et al.\ recommend to evaluate parametrizable obfuscations and randomized obfuscations on a sufficient number of parameters and random seeds. This is covered by the protection configuration sensitivity criterion $Se_{pi}$.
\paragraph{State-of-the-art SP tools} De Sutter et al.\ recommend to use state-of-the-art SP tools (such as Tigress~\cite{tigress2023} to deploy protections on samples, if possible, rather than older, outdated tools such as OLLVM~\cite{ollvm}. This recommendation is captured implicitly by the protection layering relevance criterion $Re_l$: when composing a novel obfuscation with existing ones or layering them on each other, make sure the considered existing ones are state-of-the-art. 
\paragraph{Sample Complexity} De Sutter et al.'s recommendation to use sufficiently complex samples to reflect the complexity of real-world attack steps is captured by various relevance criteria as well as by the sample feature sensitivity criterion of this framework.   

\vspace{0.2cm}

Other recommendations by De Sutter et al., such as \emph{setup specificity}, \emph{evaluation tool availability}, and \emph{sample availability}, are complementary to how an evaluation is performed but relate more to easing reproducibility and tool sharing. We obviously agree with those recommendations: better reproducibility can only make it easier to interpret results correctly. 

\section{Conclusions}
\label{sec:conclusions}

First, we analyzed four different definitions / interpretations of existing criteria for assessing the strength of sofware obfuscation techniques. Based on a literature study, we formulated 29 critiques on the existing definitions and uses of practical and theoretical potency, resilience, stealth, and cost criteria. 

Next, we formulated the eight new criteria of relevance, effectiveness, robustness,
concealment, stubbornness, sensitivity, predictability, and cost, plus one poor man's alternative of effectiveness in the form of efficacy. For those eight criteria, we formulated a total of 35 subcriteria. Our primary aim with these criteria is to provide guidance to researchers on how to improve the construct, internal, external, conclusion, and instantiation validity of their experiments and reported research results. We discussed how these criteria help to make conceptual recommendations formulated over the last five years more concrete.  

\begin{acks}
The ideas in this paper resulted from research supervised by the author and conducted in the author's team. This research was supported by the Cybersecurity Research Program Flanders, and by Research Foundation - Flanders (FWO) grant 3G0E2318. 
We thank Christian Collberg, Bart Coppens, Mila Dalla Preda, and Roberto Giacobazzi for their useful feedback on drafts of this paper. 
\end{acks}

\bibliographystyle{ACM-Reference-Format}
\bibliography{references}


\begin{thebibliography}{117}


\ifx \showCODEN    \undefined \def \showCODEN     #1{\unskip}     \fi
\ifx \showISBNx    \undefined \def \showISBNx     #1{\unskip}     \fi
\ifx \showISBNxiii \undefined \def \showISBNxiii  #1{\unskip}     \fi
\ifx \showISSN     \undefined \def \showISSN      #1{\unskip}     \fi
\ifx \showLCCN     \undefined \def \showLCCN      #1{\unskip}     \fi
\ifx \shownote     \undefined \def \shownote      #1{#1}          \fi
\ifx \showarticletitle \undefined \def \showarticletitle #1{#1}   \fi
\ifx \showURL      \undefined \def \showURL       {\relax}        \fi
\providecommand\bibfield[2]{#2}
\providecommand\bibinfo[2]{#2}
\providecommand\natexlab[1]{#1}
\providecommand\showeprint[2][]{arXiv:#2}

\bibitem[35(2024)]%
        {ninja}
\bibfield{author}{\bibinfo{person}{Vector 35}.}
  \bibinfo{year}{2024}\natexlab{}.
\newblock \bibinfo{title}{Binary Ninja}.
\newblock \bibinfo{howpublished}{\url{https://binary.ninja/}}.
\newblock


\bibitem[Abrath et~al\mbox{.}(2016)]%
        {abrath2016tightly}
\bibfield{author}{\bibinfo{person}{Bert Abrath}, \bibinfo{person}{Bart
  Coppens}, \bibinfo{person}{Stijn Volckaert}, \bibinfo{person}{Joris Wijnant},
  {and} \bibinfo{person}{Bjorn De~Sutter}.} \bibinfo{year}{2016}\natexlab{}.
\newblock \showarticletitle{Tightly-coupled self-debugging software
  protection}. In \bibinfo{booktitle}{\emph{Proc.\ 6th Workshop on Software
  Security, Protection, and Reverse Engineering}}. \bibinfo{publisher}{ACM},
  \bibinfo{address}{New York, NY, USA}, \bibinfo{pages}{1--10}.
\newblock


\bibitem[Anckaert et~al\mbox{.}(2007)]%
        {anckaert}
\bibfield{author}{\bibinfo{person}{Bertrand Anckaert}, \bibinfo{person}{Matias
  Madou}, \bibinfo{person}{Bjorn De~Sutter}, \bibinfo{person}{Bruno De~Bus},
  \bibinfo{person}{Koen De~Bosschere}, {and} \bibinfo{person}{Bart Preneel}.}
  \bibinfo{year}{2007}\natexlab{}.
\newblock \showarticletitle{Program obfuscation: a quantitative approach}. In
  \bibinfo{booktitle}{\emph{QoP '07: Proc.\ ACM workshop on Quality of
  protection}}. \bibinfo{publisher}{ACM}, \bibinfo{address}{New York, NY, USA},
  \bibinfo{pages}{15--20}.
\newblock
\href{https://doi.org/10.1145/1314257.1314263}{doi:\nolinkurl{10.1145/1314257.1314263}}


\bibitem[Bacci et~al\mbox{.}(2018)]%
        {obfuscation_detection3}
\bibfield{author}{\bibinfo{person}{Alessandro Bacci}, \bibinfo{person}{Alberto
  Bartoli}, \bibinfo{person}{Fabio Martinelli}, \bibinfo{person}{Eric Medvet},
  {and} \bibinfo{person}{Francesco Mercaldo}.} \bibinfo{year}{2018}\natexlab{}.
\newblock \showarticletitle{Detection of Obfuscation Techniques in Android
  Applications}. In \bibinfo{booktitle}{\emph{Proc.\ 13th Int'l Conf.\ on
  Availability, Reliability and Security}} \emph{(\bibinfo{series}{ARES '18})}.
  \bibinfo{publisher}{ACM}, \bibinfo{address}{New York, NY, USA}, Article
  \bibinfo{articleno}{57}, \bibinfo{numpages}{9}~pages.
\newblock
\showISBNx{9781450364485}
\href{https://doi.org/10.1145/3230833.3232823}{doi:\nolinkurl{10.1145/3230833.3232823}}


\bibitem[Banerjee et~al\mbox{.}(2021)]%
        {banerjee2021variablerecoverydecompiledbinary}
\bibfield{author}{\bibinfo{person}{Pratyay Banerjee},
  \bibinfo{person}{Kuntal~Kumar Pal}, \bibinfo{person}{Fish Wang}, {and}
  \bibinfo{person}{Chitta Baral}.} \bibinfo{year}{2021}\natexlab{}.
\newblock \bibinfo{title}{Variable Name Recovery in Decompiled Binary Code
  using Constrained Masked Language Modeling}.
\newblock
\showeprint[arxiv]{2103.12801}~[cs.LG]
\urldef\tempurl%
\url{https://arxiv.org/abs/2103.12801}
\showURL{%
\tempurl}


\bibitem[Banescu et~al\mbox{.}(2017)]%
  {2017_predicting_the_resilience_of_obfuscated_code_against_symbolic_execution_attacks_via_machine_learning}
\bibfield{author}{\bibinfo{person}{Sebastian Banescu},
  \bibinfo{person}{Christian Collberg}, {and} \bibinfo{person}{Alexander
  Pretschner}.} \bibinfo{year}{2017}\natexlab{}.
\newblock \showarticletitle{Predicting the Resilience of Obfuscated Code
  against Symbolic Execution Attacks via Machine Learning}. In
  \bibinfo{booktitle}{\emph{Proc.\ 26th USENIX Security Symposium}}
  \emph{(\bibinfo{series}{SEC'17})}. \bibinfo{publisher}{USENIX Association},
  \bibinfo{address}{USA}, \bibinfo{pages}{661--678}.
\newblock
\showISBNx{9781931971409}


\bibitem[Banescu et~al\mbox{.}(2015)]%
        {banescu15}
\bibfield{author}{\bibinfo{person}{Sebastian Banescu}, \bibinfo{person}{Martin
  Ochoa}, {and} \bibinfo{person}{Alexander Pretschner}.}
  \bibinfo{year}{2015}\natexlab{}.
\newblock \showarticletitle{A {Framework} for {Measuring} {Software}
  {Obfuscation} {Resilience} against {Automated} {Attacks}}. In
  \bibinfo{booktitle}{\emph{2015 {IEEE}/{ACM} 1st {Int'l} {Workshop} on
  {Software} {Protection}}}. \bibinfo{publisher}{IEEE},
  \bibinfo{address}{Florence, Italy}, \bibinfo{pages}{45--51}.
\newblock
\showISBNx{978-1-4673-7094-3}
\href{https://doi.org/10.1109/SPRO.2015.16}{doi:\nolinkurl{10.1109/SPRO.2015.16}}


\bibitem[Bansiya and Davis(2002)]%
        {qmood}
\bibfield{author}{\bibinfo{person}{J. Bansiya} {and} \bibinfo{person}{C.G.
  Davis}.} \bibinfo{year}{2002}\natexlab{}.
\newblock \showarticletitle{A hierarchical model for object-oriented design
  quality assessment}.
\newblock \bibinfo{journal}{\emph{IEEE Transactions on Software Engineering}}
  \bibinfo{volume}{28}, \bibinfo{number}{1} (\bibinfo{year}{2002}),
  \bibinfo{pages}{4--17}.
\newblock
\href{https://doi.org/10.1109/32.979986}{doi:\nolinkurl{10.1109/32.979986}}


\bibitem[Basile et~al\mbox{.}(2023)]%
        {Basile23}
\bibfield{author}{\bibinfo{person}{Cataldo Basile}, \bibinfo{person}{Bjorn {De
  Sutter}}, \bibinfo{person}{Daniele Canavese}, \bibinfo{person}{Leonardo
  Regano}, {and} \bibinfo{person}{Bart Coppens}.}
  \bibinfo{year}{2023}\natexlab{}.
\newblock \showarticletitle{Design, implementation, and automation of a risk
  management approach for man-at-the-End software protection}.
\newblock \bibinfo{journal}{\emph{Computers \& Security}}
  \bibinfo{volume}{132} (\bibinfo{year}{2023}), \bibinfo{pages}{103321}.
\newblock
\showISSN{0167-4048}
\href{https://doi.org/10.1016/j.cose.2023.103321}{doi:\nolinkurl{10.1016/j.cose.2023.103321}}


\bibitem[Batchelder and Hendren(2007)]%
        {batchelder2007obfuscating}
\bibfield{author}{\bibinfo{person}{Michael Batchelder} {and}
  \bibinfo{person}{Laurie Hendren}.} \bibinfo{year}{2007}\natexlab{}.
\newblock \showarticletitle{Obfuscating Java: The Most Pain for the Least
  Gain}. In \bibinfo{booktitle}{\emph{Int'l Conf.\ on Compiler Construction}},
  Vol.~\bibinfo{volume}{4420}. \bibinfo{publisher}{Springer},
  \bibinfo{pages}{96--110}.
\newblock
\showISBNx{978-3-540-71229-9}
\href{https://doi.org/10.1007/978-3-540-71229-9_7}{doi:\nolinkurl{10.1007/978-3-540-71229-9_7}}


\bibitem[Blazytko et~al\mbox{.}(2017)]%
        {2017_syntia_synthesizing_the_semantics_of_obfuscated_code}
\bibfield{author}{\bibinfo{person}{Tim Blazytko}, \bibinfo{person}{Moritz
  Contag}, \bibinfo{person}{Cornelius Aschermann}, {and}
  \bibinfo{person}{Thorsten Holz}.} \bibinfo{year}{2017}\natexlab{}.
\newblock \showarticletitle{Syntia: Synthesizing the Semantics of Obfuscated
  Code}. In \bibinfo{booktitle}{\emph{26th {USENIX} Security Symposium}}.
  \bibinfo{publisher}{{USENIX} Association}, \bibinfo{address}{Vancouver, BC},
  \bibinfo{pages}{643--659}.
\newblock
\showISBNx{978-1-931971-40-9}


\bibitem[Brunet et~al\mbox{.}(2019)]%
        {epona}
\bibfield{author}{\bibinfo{person}{Pierrick Brunet},
  \bibinfo{person}{B\'{e}atrice Creusillet}, \bibinfo{person}{Adrien Guinet},
  {and} \bibinfo{person}{Juan~Manuel Martinez}.}
  \bibinfo{year}{2019}\natexlab{}.
\newblock \showarticletitle{Epona and the Obfuscation Paradox: Transparent for
  Users and Developers, a Pain for Reversers}. In
  \bibinfo{booktitle}{\emph{Proc.\ 3rd ACM Workshop on Software Protection}}.
  \bibinfo{publisher}{ACM}, \bibinfo{address}{New York, NY, USA},
  \bibinfo{pages}{41--52}.
\newblock
\showISBNx{9781450368353}
\href{https://doi.org/10.1145/3338503.3357722}{doi:\nolinkurl{10.1145/3338503.3357722}}


\bibitem[Bruni et~al\mbox{.}(2018a)]%
        {bruni2018code2}
\bibfield{author}{\bibinfo{person}{Robert Bruni}, \bibinfo{person}{Roberto
  Giacobazzi}, {and} \bibinfo{person}{Roberta Gori}.}
  \bibinfo{year}{2018}\natexlab{a}.
\newblock \showarticletitle{Code Obfuscation Against Abstract Model Checking
  Attacks}. In \bibinfo{booktitle}{\emph{Verification, Model Checking, and
  Abstract Interpretation}}. \bibinfo{publisher}{Springer Int'l Publishing},
  \bibinfo{address}{Cham}, \bibinfo{pages}{94--115}.
\newblock
\showISBNx{978-3-319-73721-8}


\bibitem[Bruni et~al\mbox{.}(2018b)]%
        {bruni2018code1}
\bibfield{author}{\bibinfo{person}{Roberto Bruni}, \bibinfo{person}{Roberto
  Giacobazzi}, {and} \bibinfo{person}{Roberta Gori}.}
  \bibinfo{year}{2018}\natexlab{b}.
\newblock \showarticletitle{Code obfuscation against abstraction refinement
  attacks}.
\newblock \bibinfo{journal}{\emph{Formal Aspects of Computing}}
  \bibinfo{volume}{30}, \bibinfo{number}{6} (\bibinfo{year}{2018}),
  \bibinfo{pages}{685--711}.
\newblock


\bibitem[Bruni et~al\mbox{.}(2022)]%
        {bruni2022repair}
\bibfield{author}{\bibinfo{person}{Roberto Bruni}, \bibinfo{person}{Roberto
  Giacobazzi}, \bibinfo{person}{Roberta Gori}, {and} \bibinfo{person}{Francesco
  Ranzato}.} \bibinfo{year}{2022}\natexlab{}.
\newblock \showarticletitle{Abstract interpretation repair}. In
  \bibinfo{booktitle}{\emph{Proc.\ 43rd ACM SIGPLAN Int'l Conf.\ on Programming
  Language Design and Implementation}} \emph{(\bibinfo{series}{PLDI 2022})}.
  \bibinfo{publisher}{ACM}, \bibinfo{address}{New York, NY, USA},
  \bibinfo{pages}{426–441}.
\newblock
\showISBNx{9781450392655}
\href{https://doi.org/10.1145/3519939.3523453}{doi:\nolinkurl{10.1145/3519939.3523453}}


\bibitem[Callahan et~al\mbox{.}(1986)]%
        {interp_cp}
\bibfield{author}{\bibinfo{person}{David Callahan}, \bibinfo{person}{Keith~D
  Cooper}, \bibinfo{person}{Ken Kennedy}, {and} \bibinfo{person}{Linda
  Torczon}.} \bibinfo{year}{1986}\natexlab{}.
\newblock \showarticletitle{Interprocedural constant propagation}.
\newblock \bibinfo{journal}{\emph{ACM SIGPLAN Notices}} \bibinfo{volume}{21},
  \bibinfo{number}{7} (\bibinfo{year}{1986}), \bibinfo{pages}{152--161}.
\newblock


\bibitem[Campion et~al\mbox{.}(2022)]%
        {Campion2022}
\bibfield{author}{\bibinfo{person}{Marco Campion}, \bibinfo{person}{Mila
  Dalla~Preda}, {and} \bibinfo{person}{Roberto Giacobazzi}.}
  \bibinfo{year}{2022}\natexlab{}.
\newblock \showarticletitle{Partial (In)Completeness in abstract
  interpretation: limiting the imprecision in program analysis}.
\newblock \bibinfo{journal}{\emph{Proc. ACM Program. Lang.}}
  \bibinfo{volume}{6}, \bibinfo{number}{POPL}, Article \bibinfo{articleno}{59}
  (\bibinfo{date}{Jan.} \bibinfo{year}{2022}), \bibinfo{numpages}{31}~pages.
\newblock
\href{https://doi.org/10.1145/3498721}{doi:\nolinkurl{10.1145/3498721}}


\bibitem[Campion et~al\mbox{.}(2023)]%
        {campion23}
\bibfield{author}{\bibinfo{person}{Marco Campion}, \bibinfo{person}{Caterina
  Urban}, \bibinfo{person}{Mila Dalla~Preda}, {and} \bibinfo{person}{Roberto
  Giacobazzi}.} \bibinfo{year}{2023}\natexlab{}.
\newblock \showarticletitle{A Formal Framework to Measure the Incompleteness
  of Abstract Interpretations}. In \bibinfo{booktitle}{\emph{Static
  Analysis}}. \bibinfo{publisher}{Springer Nature Switzerland},
  \bibinfo{address}{Cham}, \bibinfo{pages}{114--138}.
\newblock
\showISBNx{978-3-031-44245-2}


\bibitem[Canavese et~al\mbox{.}(2017)]%
        {reganoMetric}
\bibfield{author}{\bibinfo{person}{Daniele Canavese}, \bibinfo{person}{Leonardo
  Regano}, \bibinfo{person}{Cataldo Basile}, {and} \bibinfo{person}{Alessio
  Viticchi{\'e}}.} \bibinfo{year}{2017}\natexlab{}.
\newblock \showarticletitle{Estimating Software Obfuscation Potency with
  Artificial Neural Networks}. In \bibinfo{booktitle}{\emph{Security and Trust
  Management}}, \bibfield{editor}{\bibinfo{person}{Giovanni Livraga} {and}
  \bibinfo{person}{Chris Mitchell}} (Eds.). \bibinfo{publisher}{Springer Int'l
  Publishing}, \bibinfo{address}{Cham}, \bibinfo{pages}{193--202}.
\newblock
\showISBNx{978-3-319-68063-7}


\bibitem[Ceccato(2016)]%
        {D4.06}
\bibfield{author}{\bibinfo{person}{Mariano Ceccato}.}
  \bibinfo{year}{2016}\natexlab{}.
\newblock \bibinfo{booktitle}{\emph{{ASPIRE Security Evaluation Methodology}}}.
\newblock \bibinfo{type}{Deliverable} D4.06. \bibinfo{institution}{{ASPIRE EU
  FP7 Project}}.
\newblock


\bibitem[Ceccato et~al\mbox{.}(2014)]%
        {2014afamily}
\bibfield{author}{\bibinfo{person}{Mariano Ceccato},
  \bibinfo{person}{Massimiliano Di~Penta}, \bibinfo{person}{Paolo Falcarin},
  \bibinfo{person}{Filippo Ricca}, \bibinfo{person}{Marco Torchiano}, {and}
  \bibinfo{person}{Paolo Tonella}.} \bibinfo{year}{2014}\natexlab{}.
\newblock \showarticletitle{A family of experiments to assess the effectiveness
  and efficiency of source code obfuscation techniques}.
\newblock \bibinfo{journal}{\emph{Empirical Software Engineering}}
  \bibinfo{volume}{19}, \bibinfo{number}{4} (\bibinfo{year}{2014}),
  \bibinfo{pages}{1040--1074}.
\newblock


\bibitem[Ceccato et~al\mbox{.}(2019)]%
        {emse2019}
\bibfield{author}{\bibinfo{person}{Mariano Ceccato}, \bibinfo{person}{Paolo
  Tonella}, \bibinfo{person}{Cataldo Basile}, \bibinfo{person}{Paolo Falcarin},
  \bibinfo{person}{Marco Torchiano}, \bibinfo{person}{Bart Coppens}, {and}
  \bibinfo{person}{Bjorn De~Sutter}.} \bibinfo{year}{2019}\natexlab{}.
\newblock \showarticletitle{Understanding the behaviour of hackers while
  performing attack tasks in a professional setting and in a public challenge}.
\newblock \bibinfo{journal}{\emph{Empirical Software Engineering}}
  \bibinfo{volume}{24}, \bibinfo{number}{1} (\bibinfo{year}{2019}),
  \bibinfo{pages}{240--286}.
\newblock
\showISSN{1382-3256}


\bibitem[Chidamber and Kemerer(1994)]%
        {OO-metric}
\bibfield{author}{\bibinfo{person}{S.R. Chidamber} {and} \bibinfo{person}{C.F.
  Kemerer}.} \bibinfo{year}{1994}\natexlab{}.
\newblock \showarticletitle{A metrics suite for object oriented design}.
\newblock \bibinfo{journal}{\emph{IEEE Transactions on Software Engineering}}
  \bibinfo{volume}{20}, \bibinfo{number}{6} (\bibinfo{year}{1994}),
  \bibinfo{pages}{476--493}.
\newblock
\href{https://doi.org/10.1109/32.295895}{doi:\nolinkurl{10.1109/32.295895}}


\bibitem[{Christian Collberg}(2023)]%
        {tigress2023}
\bibfield{author}{\bibinfo{person}{{Christian Collberg}}.}
  \bibinfo{year}{2023}\natexlab{}.
\newblock \bibinfo{title}{The {Tigress} {C} Obfuscator}.
\newblock
\newblock
\shownote{\url{https://tigress.wtf/}}.


\bibitem[Clarke et~al\mbox{.}(2003)]%
        {cegar}
\bibfield{author}{\bibinfo{person}{Edmund Clarke}, \bibinfo{person}{Orna
  Grumberg}, \bibinfo{person}{Somesh Jha}, \bibinfo{person}{Yuan Lu}, {and}
  \bibinfo{person}{Helmut Veith}.} \bibinfo{year}{2003}\natexlab{}.
\newblock \showarticletitle{Counterexample-guided abstraction refinement for
  symbolic model checking}.
\newblock \bibinfo{journal}{\emph{J. ACM}} \bibinfo{volume}{50},
  \bibinfo{number}{5} (\bibinfo{date}{Sept.} \bibinfo{year}{2003}),
  \bibinfo{pages}{752–794}.
\newblock
\showISSN{0004-5411}
\href{https://doi.org/10.1145/876638.876643}{doi:\nolinkurl{10.1145/876638.876643}}


\bibitem[Collberg(2024)]%
        {recipes}
\bibfield{author}{\bibinfo{person}{Christian Collberg}.}
  \bibinfo{year}{2024}\natexlab{}.
\newblock \bibinfo{title}{Tigress Recipes}.
\newblock \bibinfo{howpublished}{\url{https://tigress.wtf/recipes.html}}.
\newblock


\bibitem[Collberg et~al\mbox{.}(1997)]%
        {collberg1997taxonomy}
\bibfield{author}{\bibinfo{person}{Christian Collberg}, \bibinfo{person}{C.
  Thomborson}, {and} \bibinfo{person}{Douglas Low}.}
  \bibinfo{year}{1997}\natexlab{}.
\newblock \bibinfo{booktitle}{\emph{A Taxonomy of Obfuscating
  Transformations}}.
\newblock \bibinfo{type}{{T}echnical {R}eport} 148.
  \bibinfo{institution}{University of Auckland}.
\newblock
\urldef\tempurl%
\url{https://researchspace.auckland.ac.nz/handle/2292/3491}
\showURL{%
\tempurl}


\bibitem[Collberg et~al\mbox{.}(1998)]%
        {collberg1998manufacturing}
\bibfield{author}{\bibinfo{person}{Christian Collberg}, \bibinfo{person}{Clark
  Thomborson}, {and} \bibinfo{person}{Douglas Low}.}
  \bibinfo{year}{1998}\natexlab{}.
\newblock \showarticletitle{Manufacturing Cheap, Resilient, and Stealthy Opaque
  Constructs}. In \bibinfo{booktitle}{\emph{Proc.\ 25th ACM SIGPLAN-SIGACT
  Symposium on Principles of Programming Languages}}
  \emph{(\bibinfo{series}{POPL '98})}. \bibinfo{publisher}{ACM},
  \bibinfo{address}{New York, NY, USA}, \bibinfo{pages}{184--196}.
\newblock
\showISBNx{0897919793}
\href{https://doi.org/10.1145/268946.268962}{doi:\nolinkurl{10.1145/268946.268962}}


\bibitem[Coppens et~al\mbox{.}(2013)]%
        {coppens2013feedback}
\bibfield{author}{\bibinfo{person}{Bart Coppens}, \bibinfo{person}{Bjorn
  De~Sutter}, {and} \bibinfo{person}{Jonas Maebe}.}
  \bibinfo{year}{2013}\natexlab{}.
\newblock \showarticletitle{Feedback-driven binary code diversification}.
\newblock \bibinfo{journal}{\emph{ACM Transactions on Architecture and Code
  Optimization (TACO)}} \bibinfo{volume}{9}, \bibinfo{number}{4}
  (\bibinfo{year}{2013}), \bibinfo{pages}{1--26}.
\newblock


\bibitem[Cousot(2021)]%
        {cousot2021principles}
\bibfield{author}{\bibinfo{person}{Patrick Cousot}.}
  \bibinfo{year}{2021}\natexlab{}.
\newblock \bibinfo{booktitle}{\emph{Principles of abstract interpretation}}.
\newblock \bibinfo{publisher}{MIT Press}.
\newblock


\bibitem[Cousot and Cousot(1977)]%
        {cousot1977abstract}
\bibfield{author}{\bibinfo{person}{Patrick Cousot} {and}
  \bibinfo{person}{Radhia Cousot}.} \bibinfo{year}{1977}\natexlab{}.
\newblock \showarticletitle{Abstract interpretation: a unified lattice model
  for static analysis of programs by construction or approximation of
  fixpoints}. In \bibinfo{booktitle}{\emph{Proc.\ 4th ACM SIGACT-SIGPLAN
  symposium on Principles of programming languages}}. \bibinfo{publisher}{ACM},
  \bibinfo{address}{New York, NY, USA}, \bibinfo{pages}{238--252}.
\newblock
\href{https://doi.org/10.1145/512950.512973}{doi:\nolinkurl{10.1145/512950.512973}}


\bibitem[Cozza et~al\mbox{.}(2024)]%
        {mila24}
\bibfield{author}{\bibinfo{person}{Vittoria Cozza}, \bibinfo{person}{Mila
  {Dalla Preda}}, \bibinfo{person}{Ruggero Lanotte}, \bibinfo{person}{Marco
  Lucchesea}, \bibinfo{person}{Massimo Merroa}, {and} \bibinfo{person}{Nicola
  Zannone}.} \bibinfo{year}{2024}\natexlab{}.
\newblock \showarticletitle{Obfuscation strategies for industrial control
  systems}.
\newblock \bibinfo{journal}{\emph{Int'l Journal of Critical Infrastructure
  Protection}}  \bibinfo{volume}{47} (\bibinfo{year}{2024}),
  \bibinfo{pages}{100717}.
\newblock
\href{https://doi.org/10.1016/j.ijcip.2024.100717}{doi:\nolinkurl{10.1016/j.ijcip.2024.100717}}


\bibitem[Curtis et~al\mbox{.}(1979)]%
        {code_metrics}
\bibfield{author}{\bibinfo{person}{B. Curtis}, \bibinfo{person}{S.B. Sheppard},
  \bibinfo{person}{P. Milliman}, \bibinfo{person}{M.A. Borst}, {and}
  \bibinfo{person}{T. Love}.} \bibinfo{year}{1979}\natexlab{}.
\newblock \showarticletitle{Measuring the Psychological Complexity of Software
  Maintenance Tasks with the Halstead and McCabe Metrics}.
\newblock \bibinfo{journal}{\emph{IEEE Transactions on Software Engineering}}
  \bibinfo{volume}{SE-5}, \bibinfo{number}{2} (\bibinfo{year}{1979}),
  \bibinfo{pages}{96--104}.
\newblock
\href{https://doi.org/10.1109/TSE.1979.234165}{doi:\nolinkurl{10.1109/TSE.1979.234165}}


\bibitem[Dalla~Preda(2007)]%
        {mila07}
\bibfield{author}{\bibinfo{person}{Mila Dalla~Preda}.}
  \bibinfo{year}{2007}\natexlab{}.
\newblock \emph{\bibinfo{title}{Code obfuscation and malware detection by
  abstract interpretation}}.
\newblock \bibinfo{thesistype}{Ph.\,D. Dissertation}.
  \bibinfo{school}{Universit{\`a} degli Studi di Verona}.
\newblock


\bibitem[Dalla~Preda and Giacobazzi(2005a)]%
        {mila05b}
\bibfield{author}{\bibinfo{person}{M. Dalla~Preda} {and} \bibinfo{person}{R.
  Giacobazzi}.} \bibinfo{year}{2005}\natexlab{a}.
\newblock \showarticletitle{Control code obfuscation by abstract
  interpretation}. In \bibinfo{booktitle}{\emph{Third IEEE Int'l Conf.\ on
  Software Engineering and Formal Methods (SEFM'05)}}. \bibinfo{publisher}{IEEE
  Computer Society}, \bibinfo{address}{Los Alamitos, CA},
  \bibinfo{pages}{301--310}.
\newblock
\href{https://doi.org/10.1109/SEFM.2005.13}{doi:\nolinkurl{10.1109/SEFM.2005.13}}


\bibitem[Dalla~Preda and Giacobazzi(2005b)]%
        {mila05a}
\bibfield{author}{\bibinfo{person}{Mila Dalla~Preda} {and}
  \bibinfo{person}{Roberto Giacobazzi}.} \bibinfo{year}{2005}\natexlab{b}.
\newblock \showarticletitle{Semantic-Based Code Obfuscation by Abstract
  Interpretation}. In \bibinfo{booktitle}{\emph{Automata, Languages and
  Programming}}. \bibinfo{publisher}{Springer Berlin Heidelberg},
  \bibinfo{address}{Berlin, Heidelberg}, \bibinfo{pages}{1325--1336}.
\newblock


\bibitem[Dalla~Preda and Mastroeni(2018)]%
        {DP2018}
\bibfield{author}{\bibinfo{person}{Mila Dalla~Preda} {and}
  \bibinfo{person}{Isabella Mastroeni}.} \bibinfo{year}{2018}\natexlab{}.
\newblock \showarticletitle{Characterizing a property-driven obfuscation
  strategy}.
\newblock \bibinfo{journal}{\emph{Journal of Computer Security}}
  \bibinfo{volume}{26}, \bibinfo{number}{1} (\bibinfo{year}{2018}),
  \bibinfo{pages}{31--69}.
\newblock


\bibitem[Dalla~Preda et~al\mbox{.}(2013)]%
        {DP2013}
\bibfield{author}{\bibinfo{person}{Mila Dalla~Preda}, \bibinfo{person}{Isabella
  Mastroeni}, {and} \bibinfo{person}{Roberto Giacobazzi}.}
  \bibinfo{year}{2013}\natexlab{}.
\newblock \showarticletitle{A Formal Framework for Property-Driven Obfuscation
  Strategies}. In \bibinfo{booktitle}{\emph{Fundamentals of Computation
  Theory}}. \bibinfo{publisher}{Springer Berlin Heidelberg},
  \bibinfo{address}{Berlin, Heidelberg}, \bibinfo{pages}{133--144}.
\newblock
\showISBNx{978-3-642-40164-0}


\bibitem[{De Sutter} et~al\mbox{.}(2019)]%
        {Dagstuhl}
\bibfield{author}{\bibinfo{person}{Bjorn {De Sutter}},
  \bibinfo{person}{Christian Collberg}, \bibinfo{person}{Mila~Dalla Preda},
  {and} \bibinfo{person}{Brecht Wyseur}.} \bibinfo{year}{2019}\natexlab{}.
\newblock \showarticletitle{{Software Protection Decision Support and
  Evaluation Methodologies (Dagstuhl Seminar 19331)}}.
\newblock \bibinfo{journal}{\emph{Dagstuhl Reports}} \bibinfo{volume}{9},
  \bibinfo{number}{8} (\bibinfo{year}{2019}), \bibinfo{pages}{1--25}.
\newblock
\showISSN{2192-5283}
\href{https://doi.org/10.4230/DagRep.9.8.1}{doi:\nolinkurl{10.4230/DagRep.9.8.1}}


\bibitem[De~Sutter et~al\mbox{.}(2024)]%
        {desutter2024evaluation}
\bibfield{author}{\bibinfo{person}{Bjorn De~Sutter}, \bibinfo{person}{Sebastian
  Schrittwieser}, \bibinfo{person}{Bart Coppens}, {and}
  \bibinfo{person}{Patrick Kochberger}.} \bibinfo{year}{2024}\natexlab{}.
\newblock \showarticletitle{Evaluation Methodologies in Software Protection
  Research}.
\newblock \bibinfo{journal}{\emph{ACM Comput. Surv.}} \bibinfo{volume}{57},
  \bibinfo{number}{4}, Article \bibinfo{articleno}{86} (\bibinfo{date}{Dec.}
  \bibinfo{year}{2024}), \bibinfo{numpages}{41}~pages.
\newblock
\showISSN{0360-0300}
\href{https://doi.org/10.1145/3702314}{doi:\nolinkurl{10.1145/3702314}}


\bibitem[Dillig et~al\mbox{.}(2008)]%
        {path-sensitive}
\bibfield{author}{\bibinfo{person}{Isil Dillig}, \bibinfo{person}{Thomas
  Dillig}, {and} \bibinfo{person}{Alex Aiken}.}
  \bibinfo{year}{2008}\natexlab{}.
\newblock \showarticletitle{Sound, complete and scalable path-sensitive
  analysis}.
\newblock \bibinfo{journal}{\emph{SIGPLAN Not.}} \bibinfo{volume}{43},
  \bibinfo{number}{6} (\bibinfo{date}{June} \bibinfo{year}{2008}),
  \bibinfo{pages}{270–280}.
\newblock
\showISSN{0362-1340}
\href{https://doi.org/10.1145/1379022.1375615}{doi:\nolinkurl{10.1145/1379022.1375615}}


\bibitem[Ernst et~al\mbox{.}(2007)]%
        {Daikon}
\bibfield{author}{\bibinfo{person}{Michael~D. Ernst}, \bibinfo{person}{Jeff~H.
  Perkins}, \bibinfo{person}{Philip~J. Guo}, \bibinfo{person}{Stephen
  McCamant}, \bibinfo{person}{Carlos Pacheco}, \bibinfo{person}{Matthew~S.
  Tschantz}, {and} \bibinfo{person}{Chen Xiao}.}
  \bibinfo{year}{2007}\natexlab{}.
\newblock \showarticletitle{The Daikon system for dynamic detection of likely
  invariants}.
\newblock \bibinfo{journal}{\emph{Science of Computer Programming}}
  \bibinfo{volume}{69}, \bibinfo{number}{1} (\bibinfo{year}{2007}),
  \bibinfo{pages}{35--45}.
\newblock
\showISSN{0167-6423}
\href{https://doi.org/10.1016/j.scico.2007.01.015}{doi:\nolinkurl{10.1016/j.scico.2007.01.015}}
\newblock
\shownote{Special issue on Experimental Software and Toolkits}.


\bibitem[Faingnaert et~al\mbox{.}(2024a)]%
        {khunt++}
\bibfield{author}{\bibinfo{person}{Thomas Faingnaert}, \bibinfo{person}{Willem
  Van~Iseghem}, {and} \bibinfo{person}{Bjorn De~Sutter}.}
  \bibinfo{year}{2024}\natexlab{a}.
\newblock \showarticletitle{K-Hunt++: Improved Dynamic Cryptographic Key
  Extraction}. In \bibinfo{booktitle}{\emph{Proc.\ Workshop on Research on
  Offensive and Defensive Techniques in the Context of Man At The End (MATE)
  Attacks}} \emph{(\bibinfo{series}{CheckMATE '24})}. \bibinfo{publisher}{ACM},
  \bibinfo{address}{New York, NY, USA}, \bibinfo{pages}{22–29}.
\newblock
\showISBNx{9798400712302}
\href{https://doi.org/10.1145/3689934.3690818}{doi:\nolinkurl{10.1145/3689934.3690818}}


\bibitem[Faingnaert et~al\mbox{.}(2024b)]%
        {checkmate24}
\bibfield{author}{\bibinfo{person}{Thomas Faingnaert}, \bibinfo{person}{Tab
  Zhang}, \bibinfo{person}{Willem Van~Iseghem}, \bibinfo{person}{Gertjan
  Everaert}, \bibinfo{person}{Bart Coppens}, \bibinfo{person}{Christian
  Collberg}, {and} \bibinfo{person}{Bjorn De~Sutter}.}
  \bibinfo{year}{2024}\natexlab{b}.
\newblock \showarticletitle{Tools and Models for Software Reverse Engineering
  Research}. In \bibinfo{booktitle}{\emph{Proc.\ Workshop on Research on
  Offensive and Defensive Techniques in the Context of Man At The End (MATE)
  Attacks}} \emph{(\bibinfo{series}{CheckMATE '24})}. \bibinfo{publisher}{ACM},
  \bibinfo{address}{New York, NY, USA}, \bibinfo{pages}{44–58}.
\newblock
\showISBNx{9798400712302}
\href{https://doi.org/10.1145/3689934.3690817}{doi:\nolinkurl{10.1145/3689934.3690817}}


\bibitem[Feitelson(2023)]%
        {feitelson}
\bibfield{author}{\bibinfo{person}{Dror~G. Feitelson}.}
  \bibinfo{year}{2023}\natexlab{}.
\newblock \showarticletitle{From Code Complexity Metrics to Program
  Comprehension}.
\newblock \bibinfo{journal}{\emph{Commun. ACM}} \bibinfo{volume}{66},
  \bibinfo{number}{5} (\bibinfo{date}{April} \bibinfo{year}{2023}),
  \bibinfo{pages}{52–61}.
\newblock
\showISSN{0001-0782}
\href{https://doi.org/10.1145/3546576}{doi:\nolinkurl{10.1145/3546576}}


\bibitem[Foket et~al\mbox{.}(2014)]%
        {Foket14}
\bibfield{author}{\bibinfo{person}{Christophe Foket}, \bibinfo{person}{Bjorn
  De~Sutter}, {and} \bibinfo{person}{K. De~Bosschere}.}
  \bibinfo{year}{2014}\natexlab{}.
\newblock \showarticletitle{Pushing {Java} Type Obfuscation to the Limit}.
\newblock \bibinfo{journal}{\emph{IEEE Trans.\ on Dependable and Secure
  Computing}} \bibinfo{volume}{11}, \bibinfo{number}{6} (\bibinfo{date}{2}
  \bibinfo{year}{2014}), \bibinfo{pages}{553--567}.
\newblock


\bibitem[{Gartner, Inc.}(2019)]%
        {Gartner-report-riskanalysis}
\bibfield{author}{\bibinfo{person}{{Gartner, Inc.}}}
  \bibinfo{year}{2019}\natexlab{}.
\newblock \bibinfo{title}{Risk Assessment Process and Methodologies Primer for
  2019}.
\newblock \bibinfo{howpublished}{Online at
  \url{https://www.gartner.com/en/documents/3938592}}.
\newblock


\bibitem[Giacobazzi(2008)]%
        {2008hiding}
\bibfield{author}{\bibinfo{person}{Roberto Giacobazzi}.}
  \bibinfo{year}{2008}\natexlab{}.
\newblock \showarticletitle{Hiding Information in Completeness Holes: New
  Perspectives in Code Obfuscation and Watermarking}. In
  \bibinfo{booktitle}{\emph{Proc.\ Sixth IEEE Int'l Conf.\ on Software
  Engineering and Formal Methods}} \emph{(\bibinfo{series}{SEFM '08})}.
  \bibinfo{publisher}{IEEE Computer Society}, \bibinfo{address}{USA},
  \bibinfo{pages}{7–18}.
\newblock
\showISBNx{9780769534374}
\href{https://doi.org/10.1109/SEFM.2008.41}{doi:\nolinkurl{10.1109/SEFM.2008.41}}


\bibitem[Giacobazzi et~al\mbox{.}(2012)]%
        {Roberto2012}
\bibfield{author}{\bibinfo{person}{Roberto Giacobazzi}, \bibinfo{person}{Neil
  Jones}, {and} \bibinfo{person}{Isabella Mastroeni}.}
  \bibinfo{year}{2012}\natexlab{}.
\newblock \showarticletitle{Obfuscation by partial evaluation of distorted
  interpreters}. In \bibinfo{booktitle}{\emph{Conf.\ Record of the Annual ACM
  Symposium on Principles of Programming Languages}}. \bibinfo{publisher}{ACM},
  \bibinfo{address}{New York, NY, USA}, \bibinfo{pages}{63--72}.
\newblock
\href{https://doi.org/10.1145/2103746.2103761}{doi:\nolinkurl{10.1145/2103746.2103761}}


\bibitem[Giacobazzi and Mastroeni(2012)]%
        {2012_making}
\bibfield{author}{\bibinfo{person}{Roberto Giacobazzi} {and}
  \bibinfo{person}{Isabella Mastroeni}.} \bibinfo{year}{2012}\natexlab{}.
\newblock \showarticletitle{Making Abstract Interpretation Incomplete: Modeling
  the Potency of Obfuscation}. In \bibinfo{booktitle}{\emph{Static Analysis}}.
  \bibinfo{publisher}{Springer Berlin Heidelberg}, \bibinfo{address}{Berlin,
  Heidelberg}, \bibinfo{pages}{129--145}.
\newblock
\showISBNx{978-3-642-33125-1}


\bibitem[Giacobazzi et~al\mbox{.}(2017)]%
        {2017GMDP}
\bibfield{author}{\bibinfo{person}{Roberto Giacobazzi},
  \bibinfo{person}{Isabella Mastroeni}, {and} \bibinfo{person}{Mila
  Dalla~Preda}.} \bibinfo{year}{2017}\natexlab{}.
\newblock \showarticletitle{Maximal incompleteness as obfuscation potency}.
\newblock \bibinfo{journal}{\emph{Formal Aspects of Computing}}
  \bibinfo{volume}{29}, \bibinfo{number}{1} (\bibinfo{year}{2017}),
  \bibinfo{pages}{3--31}.
\newblock


\bibitem[Giacobazzi et~al\mbox{.}(2023)]%
        {fitting_roberto}
\bibfield{author}{\bibinfo{person}{Roberto Giacobazzi},
  \bibinfo{person}{Isabella Mastroeni}, {and} \bibinfo{person}{Elia
  Perantoni}.} \bibinfo{year}{2023}\natexlab{}.
\newblock \showarticletitle{How Fitting is Your Abstract Domain?}. In
  \bibinfo{booktitle}{\emph{Static Analysis}}. \bibinfo{publisher}{Springer
  Nature Switzerland}, \bibinfo{address}{Cham}, \bibinfo{pages}{286--309}.
\newblock
\showISBNx{978-3-031-44245-2}


\bibitem[Giacobazzi and Ranzato(2025)]%
        {2025roberto}
\bibfield{author}{\bibinfo{person}{Roberto Giacobazzi} {and}
  \bibinfo{person}{Francesco Ranzato}.} \bibinfo{year}{2025}\natexlab{}.
\newblock \showarticletitle{The Best of Abstract Interpretations}.
\newblock \bibinfo{journal}{\emph{Proc. ACM Program. Lang.}}
  \bibinfo{volume}{9}, \bibinfo{number}{POPL}, Article \bibinfo{articleno}{46}
  (\bibinfo{date}{Jan.} \bibinfo{year}{2025}), \bibinfo{numpages}{31}~pages.
\newblock
\href{https://doi.org/10.1145/3704882}{doi:\nolinkurl{10.1145/3704882}}


\bibitem[Giacobazzi and Toppan(2012)]%
        {entropy_metric}
\bibfield{author}{\bibinfo{person}{Roberto Giacobazzi} {and}
  \bibinfo{person}{Andrea Toppan}.} \bibinfo{year}{2012}\natexlab{}.
\newblock \showarticletitle{On Entropy Measures for Code Obfuscation}. In
  \bibinfo{booktitle}{\emph{Proc.\ ACM SIGPLAN Software Security and Protection
  Workshop}}.
\newblock


\bibitem[Gilray et~al\mbox{.}(2016)]%
        {polyvariance}
\bibfield{author}{\bibinfo{person}{Thomas Gilray}, \bibinfo{person}{Michael~D.
  Adams}, {and} \bibinfo{person}{Matthew Might}.}
  \bibinfo{year}{2016}\natexlab{}.
\newblock \showarticletitle{Allocation characterizes polyvariance: a unified
  methodology for polyvariant control-flow analysis}.
\newblock \bibinfo{journal}{\emph{SIGPLAN Not.}} \bibinfo{volume}{51},
  \bibinfo{number}{9} (\bibinfo{date}{Sept.} \bibinfo{year}{2016}),
  \bibinfo{pages}{407–420}.
\newblock
\showISSN{0362-1340}
\href{https://doi.org/10.1145/3022670.2951936}{doi:\nolinkurl{10.1145/3022670.2951936}}


\bibitem[{GNU Project}(2024)]%
        {binutils}
\bibfield{author}{\bibinfo{person}{{GNU Project}}.}
  \bibinfo{year}{2024}\natexlab{}.
\newblock \bibinfo{title}{objdump - {GNU Binutils}}.
\newblock \bibinfo{howpublished}{\url{https://www.gnu.org/software/binutils/}}.
\newblock


\bibitem[Grove and Chambers(2001)]%
        {vortex}
\bibfield{author}{\bibinfo{person}{David Grove} {and} \bibinfo{person}{Craig
  Chambers}.} \bibinfo{year}{2001}\natexlab{}.
\newblock \showarticletitle{A framework for call graph construction
  algorithms}.
\newblock \bibinfo{journal}{\emph{ACM Trans. Program. Lang. Syst.}}
  \bibinfo{volume}{23}, \bibinfo{number}{6} (\bibinfo{date}{Nov.}
  \bibinfo{year}{2001}), \bibinfo{pages}{685–746}.
\newblock
\showISSN{0164-0925}
\href{https://doi.org/10.1145/506315.506316}{doi:\nolinkurl{10.1145/506315.506316}}


\bibitem[Guelton et~al\mbox{.}(2018)]%
        {optimization}
\bibfield{author}{\bibinfo{person}{S. Guelton}, \bibinfo{person}{A. Guinet},
  \bibinfo{person}{P. Brunet}, \bibinfo{person}{J.~M. Martinez},
  \bibinfo{person}{F. Dagnat}, {and} \bibinfo{person}{N. Szlifierski}.}
  \bibinfo{year}{2018}\natexlab{}.
\newblock \showarticletitle{Combining Obfuscation and Optimizations in the Real
  World}. In \bibinfo{booktitle}{\emph{IEEE 18th Int'l Working Conf.\ on Source
  Code Analysis and Manipulation (SCAM)}}. \bibinfo{pages}{24--33}.
\newblock
\showISSN{2470-6892}
\href{https://doi.org/10.1109/SCAM.2018.00010}{doi:\nolinkurl{10.1109/SCAM.2018.00010}}


\bibitem[Halstead(1977)]%
        {halstead}
\bibfield{author}{\bibinfo{person}{Maurice~H Halstead}.}
  \bibinfo{year}{1977}\natexlab{}.
\newblock \bibinfo{booktitle}{\emph{Elements of Software Science (Operating and
  programming systems series)}}.
\newblock \bibinfo{publisher}{Elsevier Science Inc.}
\newblock


\bibitem[H{\"a}nsch et~al\mbox{.}(2018)]%
  {2018_programming_experience_might_not_help_in_comprehending_obfuscated_source_code_efficiently}
\bibfield{author}{\bibinfo{person}{Norman H{\"a}nsch}, \bibinfo{person}{Andrea
  Schankin}, \bibinfo{person}{Mykolai Protsenko}, \bibinfo{person}{Felix
  Freiling}, {and} \bibinfo{person}{Zinaida Benenson}.}
  \bibinfo{year}{2018}\natexlab{}.
\newblock \showarticletitle{Programming Experience Might Not Help in
  Comprehending Obfuscated Source Code Efficiently}. In
  \bibinfo{booktitle}{\emph{SOUPS}}. \bibinfo{publisher}{{USENIX} Association},
  \bibinfo{address}{Baltimore, MD}, \bibinfo{pages}{341--356}.
\newblock
\showISBNx{978-1-939133-10-6}


\bibitem[Harrison and Magel(1981)]%
        {harrison1981}
\bibfield{author}{\bibinfo{person}{Warren~A Harrison} {and}
  \bibinfo{person}{Kenneth~I Magel}.} \bibinfo{year}{1981}\natexlab{}.
\newblock \showarticletitle{A complexity measure based on nesting level}.
\newblock \bibinfo{journal}{\emph{ACM Sigplan Notices}} \bibinfo{volume}{16},
  \bibinfo{number}{3} (\bibinfo{year}{1981}), \bibinfo{pages}{63--74}.
\newblock


\bibitem[Henry and Kafura(1981)]%
        {flow_metrics}
\bibfield{author}{\bibinfo{person}{Sallie Henry} {and} \bibinfo{person}{Dennis
  Kafura}.} \bibinfo{year}{1981}\natexlab{}.
\newblock \showarticletitle{Software structure metrics based on information
  flow}.
\newblock \bibinfo{journal}{\emph{IEEE Transactions on Software Engineering}}
  \bibinfo{volume}{SE-7}, \bibinfo{number}{5} (\bibinfo{year}{1981}),
  \bibinfo{pages}{510--518}.
\newblock


\bibitem[Hex-Rays(2024)]%
        {IDA}
\bibfield{author}{\bibinfo{person}{Hex-Rays}.} \bibinfo{year}{2024}\natexlab{}.
\newblock \bibinfo{title}{IDA Pro}.
\newblock \bibinfo{howpublished}{\url{https://hex-rays.com/ida-pro}}.
\newblock


\bibitem[Jiang et~al\mbox{.}(2021)]%
        {obfuscation_detection2}
\bibfield{author}{\bibinfo{person}{Shuai Jiang}, \bibinfo{person}{Yao Hong},
  \bibinfo{person}{Cai Fu}, \bibinfo{person}{Yekui Qian}, {and}
  \bibinfo{person}{Lansheng Han}.} \bibinfo{year}{2021}\natexlab{}.
\newblock \showarticletitle{Function-level obfuscation detection method based
  on Graph Convolutional Networks}.
\newblock \bibinfo{journal}{\emph{Journal of Information Security and
  Applications}}  \bibinfo{volume}{61} (\bibinfo{year}{2021}),
  \bibinfo{pages}{102953}.
\newblock
\showISSN{2214-2126}
\href{https://doi.org/10.1016/j.jisa.2021.102953}{doi:\nolinkurl{10.1016/j.jisa.2021.102953}}


\bibitem[{Joint Task Force Transformation Initiative}(2011)]%
        {nistSP800-39}
\bibfield{author}{\bibinfo{person}{{Joint Task Force Transformation
  Initiative}}.} \bibinfo{year}{2011}\natexlab{}.
\newblock \bibinfo{booktitle}{\emph{{SP 800-39}. Managing Information Security
  Risk: Organization, Mission, and Information System View}}.
\newblock \bibinfo{type}{{T}echnical {R}eport}. \bibinfo{institution}{National
  Institute of Standards \& Technology}.
\newblock


\bibitem[Junod et~al\mbox{.}(2015)]%
        {ollvm}
\bibfield{author}{\bibinfo{person}{P. Junod}, \bibinfo{person}{J. Rinaldini},
  \bibinfo{person}{J. Wehrli}, {and} \bibinfo{person}{J. Michielin}.}
  \bibinfo{year}{2015}\natexlab{}.
\newblock \showarticletitle{Obfuscator-LLVM -- Software Protection for the
  Masses}. In \bibinfo{booktitle}{\emph{IEEE/ACM 1st Int'l Workshop on Software
  Protection (SPRO)}}, Vol.~\bibinfo{volume}{00}. \bibinfo{pages}{3--9}.
\newblock
\href{https://doi.org/10.1109/SPRO.2015.10}{doi:\nolinkurl{10.1109/SPRO.2015.10}}


\bibitem[Kanzaki et~al\mbox{.}(2015)]%
        {code_artificiality}
\bibfield{author}{\bibinfo{person}{Y. Kanzaki}, \bibinfo{person}{A. Monden},
  {and} \bibinfo{person}{C. Collberg}.} \bibinfo{year}{2015}\natexlab{}.
\newblock \showarticletitle{Code Artificiality: A Metric for the Code Stealth
  Based on an N-Gram Model}. In \bibinfo{booktitle}{\emph{IEEE/ACM 1st Int'l
  Workshop on Software Protection}}. \bibinfo{publisher}{IEEE Press},
  \bibinfo{pages}{31--37}.
\newblock
\href{https://doi.org/10.1109/SPRO.2015.14}{doi:\nolinkurl{10.1109/SPRO.2015.14}}


\bibitem[Kim et~al\mbox{.}(2021)]%
        {obfuscation_detection6}
\bibfield{author}{\bibinfo{person}{Jeongwoo Kim}, \bibinfo{person}{Seoyeon
  Kang}, \bibinfo{person}{Eun-Sun Cho}, {and} \bibinfo{person}{Joon-Young
  Paik}.} \bibinfo{year}{2021}\natexlab{}.
\newblock \showarticletitle{LOM: Lightweight Classifier for Obfuscation
  Methods}. In \bibinfo{booktitle}{\emph{Information Security Applications}}.
  \bibinfo{publisher}{Springer Int'l Publishing}, \bibinfo{address}{Cham},
  \bibinfo{pages}{3--15}.
\newblock
\showISBNx{978-3-030-89432-0}


\bibitem[{Kinder}(2012)]%
        {kinder}
\bibfield{author}{\bibinfo{person}{J. {Kinder}}.}
  \bibinfo{year}{2012}\natexlab{}.
\newblock \showarticletitle{Towards Static Analysis of
  Virtualization-Obfuscated Binaries}. In \bibinfo{booktitle}{\emph{2012 19th
  Working Conf.\ on Reverse Engineering}}. \bibinfo{pages}{61--70}.
\newblock
\showISSN{1095-1350}
\href{https://doi.org/10.1109/WCRE.2012.16}{doi:\nolinkurl{10.1109/WCRE.2012.16}}


\bibitem[Li et~al\mbox{.}(2018)]%
        {khunt}
\bibfield{author}{\bibinfo{person}{Juanru Li}, \bibinfo{person}{Zhiqiang Lin},
  \bibinfo{person}{Juan Caballero}, \bibinfo{person}{Yuanyuan Zhang}, {and}
  \bibinfo{person}{Dawu Gu}.} \bibinfo{year}{2018}\natexlab{}.
\newblock \showarticletitle{K-{{Hunt}}: {{Pinpointing Insecure Cryptographic
  Keys}} from {{Execution Traces}}}. In \bibinfo{booktitle}{\emph{Proc.\ {{ACM
  SIGSAC Conf.}} on {{Computer}} and {{Communications Security}}}}
  (2018-10-15). \bibinfo{publisher}{ACM}, \bibinfo{address}{New York, NY, USA},
  \bibinfo{pages}{412--425}.
\newblock
\showISBNx{978-1-4503-5693-0}
\href{https://doi.org/10.1145/3243734.3243783}{doi:\nolinkurl{10.1145/3243734.3243783}}


\bibitem[Liem et~al\mbox{.}(2008)]%
        {layering}
\bibfield{author}{\bibinfo{person}{Clifford Liem}, \bibinfo{person}{Yuan~Xiang
  Gu}, {and} \bibinfo{person}{Harold Johnson}.}
  \bibinfo{year}{2008}\natexlab{}.
\newblock \showarticletitle{A Compiler-Based Infrastructure for
  Software-Protection}. In \bibinfo{booktitle}{\emph{Proc.\ Third ACM SIGPLAN
  Workshop on Programming Languages and Analysis for Security}}
  \emph{(\bibinfo{series}{PLAS '08})}. \bibinfo{publisher}{ACM},
  \bibinfo{address}{New York, NY, USA}, \bibinfo{pages}{33--44}.
\newblock
\showISBNx{9781595939364}
\href{https://doi.org/10.1145/1375696.1375702}{doi:\nolinkurl{10.1145/1375696.1375702}}


\bibitem[Linn and Debray(2003)]%
        {Linn2003}
\bibfield{author}{\bibinfo{person}{Cullen Linn} {and} \bibinfo{person}{Saumya
  Debray}.} \bibinfo{year}{2003}\natexlab{}.
\newblock \showarticletitle{{Obfuscation of executable code to improve
  resistance to static disassembly}}. In \bibinfo{booktitle}{\emph{Proc.\ 10th
  ACM Conf.\ on Computer and communications security (CCS '03)}}.
  \bibinfo{publisher}{ACM Press}, \bibinfo{address}{New York, New York, USA},
  \bibinfo{pages}{290--299}.
\newblock
\showISBNx{1581137389}
\href{https://doi.org/10.1145/948109.948149}{doi:\nolinkurl{10.1145/948109.948149}}


\bibitem[Liu et~al\mbox{.}(2021)]%
  {2021_mba_blast_unveiling_and_simplifying_mixed_boolean_arithmetic_obfuscation}
\bibfield{author}{\bibinfo{person}{Binbin Liu}, \bibinfo{person}{Junfu Shen},
  \bibinfo{person}{Jiang Ming}, \bibinfo{person}{Qilong Zheng},
  \bibinfo{person}{Jing Li}, {and} \bibinfo{person}{Dongpeng Xu}.}
  \bibinfo{year}{2021}\natexlab{}.
\newblock \showarticletitle{{MBA-Blast}: Unveiling and Simplifying Mixed
  {Boolean-Arithmetic} Obfuscation}. In \bibinfo{booktitle}{\emph{30th USENIX
  Security Symposium}}. \bibinfo{publisher}{USENIX Association},
  \bibinfo{pages}{1701--1718}.
\newblock
\showISBNx{978-1-939133-24-3}


\bibitem[Liu et~al\mbox{.}(2017)]%
        {liu2017stochastic}
\bibfield{author}{\bibinfo{person}{H. Liu}, \bibinfo{person}{C. Sun},
  \bibinfo{person}{Z. Su}, \bibinfo{person}{Y. Jiang}, \bibinfo{person}{M. Gu},
  {and} \bibinfo{person}{J. Sun}.} \bibinfo{year}{2017}\natexlab{}.
\newblock \showarticletitle{Stochastic Optimization of Program Obfuscation}. In
  \bibinfo{booktitle}{\emph{IEEE/ACM 39th Int'l Conf.\ on Software Engineering
  (ICSE)}}. \bibinfo{pages}{221--231}.
\newblock
\showISSN{1558-1225}
\href{https://doi.org/10.1109/ICSE.2017.28}{doi:\nolinkurl{10.1109/ICSE.2017.28}}


\bibitem[Lomne(2016)]%
        {common_criteria}
\bibfield{author}{\bibinfo{person}{Victor Lomne}.}
  \bibinfo{year}{2016}\natexlab{}.
\newblock \showarticletitle{Common criteria certification of a smartcard: a
  technical overview}. In \bibinfo{booktitle}{\emph{Proc. Int. Workshop
  Cryptographic Hardware Embedded Syst. Tut}}. \bibinfo{pages}{1--105}.
\newblock


\bibitem[Lukyanenko et~al\mbox{.}(2014)]%
        {lukyanenko2014instantiation}
\bibfield{author}{\bibinfo{person}{Roman Lukyanenko}, \bibinfo{person}{Joerg
  Evermann}, {and} \bibinfo{person}{Jeffrey Parsons}.}
  \bibinfo{year}{2014}\natexlab{}.
\newblock \showarticletitle{Instantiation validity in IS design research}. In
  \bibinfo{booktitle}{\emph{Prooc.\ 9th Int'l Conf.\ Advancing the Impact of
  Design Science: Moving from Theory to Practice}}. Springer,
  \bibinfo{pages}{321--328}.
\newblock


\bibitem[Marastoni et~al\mbox{.}(2021)]%
        {marastoni2021data}
\bibfield{author}{\bibinfo{person}{Niccol{\`o} Marastoni},
  \bibinfo{person}{Roberto Giacobazzi}, {and} \bibinfo{person}{Mila
  Dalla~Preda}.} \bibinfo{year}{2021}\natexlab{}.
\newblock \showarticletitle{Data augmentation and transfer learning to classify
  malware images in a deep learning context}.
\newblock \bibinfo{journal}{\emph{Journal of Computer Virology and Hacking
  Techniques}} \bibinfo{volume}{17}, \bibinfo{number}{4}
  (\bibinfo{year}{2021}), \bibinfo{pages}{279--297}.
\newblock


\bibitem[Mariano et~al\mbox{.}(2024)]%
        {mariano24}
\bibfield{author}{\bibinfo{person}{Benjamin Mariano}, \bibinfo{person}{Ziteng
  Wang}, \bibinfo{person}{Shankara Pailoor}, \bibinfo{person}{Christian
  Collberg}, {and} \bibinfo{person}{I\c{s}il Dillig}.}
  \bibinfo{year}{2024}\natexlab{}.
\newblock \showarticletitle{Control-Flow Deobfuscation using Trace-Informed
  Compositional Program Synthesis}.
\newblock \bibinfo{journal}{\emph{Proc. ACM Program. Lang.}}
  \bibinfo{volume}{8}, \bibinfo{number}{OOPSLA2}, Article
  \bibinfo{articleno}{349} (\bibinfo{date}{Oct.} \bibinfo{year}{2024}),
  \bibinfo{numpages}{31}~pages.
\newblock
\href{https://doi.org/10.1145/3689789}{doi:\nolinkurl{10.1145/3689789}}


\bibitem[McCabe(1976)]%
        {mccabe1976}
\bibfield{author}{\bibinfo{person}{Thomas~J McCabe}.}
  \bibinfo{year}{1976}\natexlab{}.
\newblock \showarticletitle{A complexity measure}.
\newblock \bibinfo{journal}{\emph{IEEE Transactions on software Engineering}}
  \bibinfo{number}{4} (\bibinfo{year}{1976}), \bibinfo{pages}{308--320}.
\newblock


\bibitem[Menguy et~al\mbox{.}(2021)]%
  {2021_search_based_local_black_box_deobfuscation_understand_improve_and_mitigate}
\bibfield{author}{\bibinfo{person}{Gr\'{e}goire Menguy},
  \bibinfo{person}{S\'{e}bastien Bardin}, \bibinfo{person}{Richard Bonichon},
  {and} \bibinfo{person}{Cauim de~Souza Lima}.}
  \bibinfo{year}{2021}\natexlab{}.
\newblock \showarticletitle{Search-Based Local Black-Box Deobfuscation:
  Understand, Improve and Mitigate}. In \bibinfo{booktitle}{\emph{Proc.\ ACM
  SIGSAC Conf.\ on Computer and Communications Security}}
  \emph{(\bibinfo{series}{CCS '21})}. \bibinfo{publisher}{ACM},
  \bibinfo{address}{New York, NY, USA}, \bibinfo{pages}{2513--2525}.
\newblock
\showISBNx{9781450384544}
\href{https://doi.org/10.1145/3460120.3485250}{doi:\nolinkurl{10.1145/3460120.3485250}}


\bibitem[Munson and Kohshgoftaar(1993)]%
        {munson1993}
\bibfield{author}{\bibinfo{person}{John~C Munson} {and}
  \bibinfo{person}{Taghi~M Kohshgoftaar}.} \bibinfo{year}{1993}\natexlab{}.
\newblock \showarticletitle{Measurement of data structure complexity}.
\newblock \bibinfo{journal}{\emph{Journal of Systems and Software}}
  \bibinfo{volume}{20}, \bibinfo{number}{3} (\bibinfo{year}{1993}),
  \bibinfo{pages}{217--225}.
\newblock


\bibitem[Nagra and Collberg(2009)]%
        {collbergbook}
\bibfield{author}{\bibinfo{person}{Jasvir Nagra} {and}
  \bibinfo{person}{Christian Collberg}.} \bibinfo{year}{2009}\natexlab{}.
\newblock \bibinfo{booktitle}{\emph{Surreptitious Software: Obfuscation,
  Watermarking, and Tamperproofing for Software Protection: Obfuscation,
  Watermarking, and Tamperproofing for Software Protection}}.
\newblock \bibinfo{publisher}{Pearson Education}.
\newblock


\bibitem[Nakamura et~al\mbox{.}(2003)]%
        {mental_metrics}
\bibfield{author}{\bibinfo{person}{M. Nakamura}, \bibinfo{person}{A. Monden},
  \bibinfo{person}{T. Itoh}, \bibinfo{person}{K. Matsumoto},
  \bibinfo{person}{Y. Kanzaki}, {and} \bibinfo{person}{H. Satoh}.}
  \bibinfo{year}{2003}\natexlab{}.
\newblock \showarticletitle{Queue-based cost evaluation of mental simulation
  process in program comprehension}. In \bibinfo{booktitle}{\emph{Proc.\ 5th
  Int'l Workshop on Enterprise Networking and Computing in Healthcare Industry
  (IEEE Cat. No.03EX717)}}. \bibinfo{publisher}{IEEE Computer Society},
  \bibinfo{pages}{351--360}.
\newblock
\href{https://doi.org/10.1109/METRIC.2003.1232480}{doi:\nolinkurl{10.1109/METRIC.2003.1232480}}


\bibitem[Ni et~al\mbox{.}(2018)]%
        {ni2018malware}
\bibfield{author}{\bibinfo{person}{Sang Ni}, \bibinfo{person}{Quan Qian}, {and}
  \bibinfo{person}{Rui Zhang}.} \bibinfo{year}{2018}\natexlab{}.
\newblock \showarticletitle{Malware identification using visualization images
  and deep learning}.
\newblock \bibinfo{journal}{\emph{Computers \& Security}}
  (\bibinfo{year}{2018}).
\newblock


\bibitem[Oviedo(1980)]%
        {oviedo80}
\bibfield{author}{\bibinfo{person}{E.I. Oviedo}.}
  \bibinfo{year}{1980}\natexlab{}.
\newblock \showarticletitle{Control Flow, Data Flow, and Program Complexity}.
  In \bibinfo{booktitle}{\emph{Proc.\ {IEEE COMPSAC}}}.
  \bibinfo{pages}{146--152}.
\newblock


\bibitem[Parker. et~al\mbox{.}(2021)]%
        {secrypt21}
\bibfield{author}{\bibinfo{person}{Colby~B. Parker.}, \bibinfo{person}{J.~Todd
  McDonald.}, {and} \bibinfo{person}{Dimitrios Damopoulos.}}
  \bibinfo{year}{2021}\natexlab{}.
\newblock \bibinfo{title}{Machine Learning Classification of Obfuscation using
  Image Visualization}.
\newblock \bibinfo{numpages}{854--859}~pages.
\newblock
\showISBNx{978-989-758-524-1}
\showISSN{2184-7711}
\href{https://doi.org/10.5220/0010607408540859}{doi:\nolinkurl{10.5220/0010607408540859}}


\bibitem[Powers(2011)]%
        {statistics}
\bibfield{author}{\bibinfo{person}{David~Martin Powers}.}
  \bibinfo{year}{2011}\natexlab{}.
\newblock \showarticletitle{From precision, recall and {F-measure} to {ROC},
  informedness, markedness and correlation}.
\newblock \bibinfo{journal}{\emph{Int. J. Mach. Learn. Technol.}}
  \bibinfo{volume}{2}, \bibinfo{number}{1} (\bibinfo{year}{2011}),
  \bibinfo{pages}{37--63}.
\newblock


\bibitem[Raubitzek et~al\mbox{.}(2024)]%
        {obfuscation_detection4}
\bibfield{author}{\bibinfo{person}{Sebastian Raubitzek},
  \bibinfo{person}{Sebastian Schrittwieser}, \bibinfo{person}{Caroline
  Lawitschka}, \bibinfo{person}{Kevin Mallinger}, \bibinfo{person}{Andreas
  Ekelhart}, {and} \bibinfo{person}{Edgar~R Weippl}.}
  \bibinfo{year}{2024}\natexlab{}.
\newblock \showarticletitle{Code Obfuscation Classification using Singular
  Value Decomposition on Grayscale Image Representations}.
\newblock In \bibinfo{booktitle}{\emph{SECRYPT}}.
\newblock


\bibitem[Regano et~al\mbox{.}(2017)]%
        {reganoL2P}
\bibfield{author}{\bibinfo{person}{Leonardo Regano}, \bibinfo{person}{Daniele
  Canavese}, \bibinfo{person}{Cataldo Basile}, {and} \bibinfo{person}{Antonio
  Lioy}.} \bibinfo{year}{2017}\natexlab{}.
\newblock \showarticletitle{Towards Optimally Hiding Protected Assets in
  Software Applications}. In \bibinfo{booktitle}{\emph{Proc.\ Int'l Conf.\ on
  Software Quality, Reliability and Security}}. \bibinfo{publisher}{IEEE
  Computer Society}, \bibinfo{pages}{374--385}.
\newblock


\bibitem[Rolles(2009)]%
        {deobf_virtualization}
\bibfield{author}{\bibinfo{person}{Rolf Rolles}.}
  \bibinfo{year}{2009}\natexlab{}.
\newblock \showarticletitle{Unpacking Virtualization Obfuscators}. In
  \bibinfo{booktitle}{\emph{Proc.\ 3rd USENIX Conf.\ on Offensive
  Technologies}} \emph{(\bibinfo{series}{WOOT'09})}. \bibinfo{publisher}{USENIX
  Association}, \bibinfo{pages}{1--7}.
\newblock


\bibitem[Schloegel et~al\mbox{.}(2022)]%
        {2022_loki_hardening_code_obfuscation_against_automated_attacks}
\bibfield{author}{\bibinfo{person}{Moritz Schloegel}, \bibinfo{person}{Tim
  Blazytko}, \bibinfo{person}{Moritz Contag}, \bibinfo{person}{Cornelius
  Aschermann}, \bibinfo{person}{Julius Basler}, \bibinfo{person}{Thorsten
  Holz}, {and} \bibinfo{person}{Ali Abbasi}.} \bibinfo{year}{2022}\natexlab{}.
\newblock \showarticletitle{Loki: Hardening Code Obfuscation Against Automated
  Attacks}. In \bibinfo{booktitle}{\emph{31st USENIX Security Symposium}}.
  \bibinfo{publisher}{USENIX Association}, \bibinfo{address}{Boston, MA},
  \bibinfo{pages}{3055--3073}.
\newblock
\showISBNx{978-1-939133-31-1}


\bibitem[Schrittwieser et~al\mbox{.}(2016)]%
        {survey2016}
\bibfield{author}{\bibinfo{person}{Sebastian Schrittwieser},
  \bibinfo{person}{Stefan Katzenbeisser}, \bibinfo{person}{Johannes Kinder},
  \bibinfo{person}{Georg Merzdovnik}, {and} \bibinfo{person}{Edgar Weippl}.}
  \bibinfo{year}{2016}\natexlab{}.
\newblock \showarticletitle{Protecting Software through Obfuscation: Can It
  Keep Pace with Progress in Code Analysis?}
\newblock \bibinfo{journal}{\emph{ACM Comput. Surv.}} \bibinfo{volume}{49},
  \bibinfo{number}{1}, Article \bibinfo{articleno}{4} (\bibinfo{date}{apr}
  \bibinfo{year}{2016}), \bibinfo{numpages}{37}~pages.
\newblock
\showISSN{0360-0300}
\href{https://doi.org/10.1145/2886012}{doi:\nolinkurl{10.1145/2886012}}


\bibitem[Schrittwieser et~al\mbox{.}(2024)]%
        {modeling_stealth}
\bibfield{author}{\bibinfo{person}{Sebastian Schrittwieser},
  \bibinfo{person}{Elisabeth Wimmer}, \bibinfo{person}{Kevin Mallinger},
  \bibinfo{person}{Patrick Kochberger}, \bibinfo{person}{Caroline Lawitschka},
  \bibinfo{person}{Sebastian Raubitzek}, {and} \bibinfo{person}{Edgar~R.
  Weippl}.} \bibinfo{year}{2024}\natexlab{}.
\newblock \showarticletitle{Modeling Obfuscation Stealth Through Code
  Complexity}. In \bibinfo{booktitle}{\emph{Computer Security. ESORICS 2023
  Int'l Workshops}}. \bibinfo{publisher}{Springer Nature Switzerland},
  \bibinfo{address}{Cham}, \bibinfo{pages}{392--408}.
\newblock
\showISBNx{978-3-031-54129-2}


\bibitem[Shijia et~al\mbox{.}(2022)]%
  {2022_chosen_instruction_attack_against_commercial_code_virtualization_obfuscators}
\bibfield{author}{\bibinfo{person}{Li Shijia}, \bibinfo{person}{Jia Chunfu},
  \bibinfo{person}{Qiu Pengda}, \bibinfo{person}{Chen Qiyuan},
  \bibinfo{person}{Ming Jiang}, {and} \bibinfo{person}{Gao Debin}.}
  \bibinfo{year}{2022}\natexlab{}.
\newblock \showarticletitle{Chosen-Instruction Attack Against Commercial Code
  Virtualization Obfuscators}. In \bibinfo{booktitle}{\emph{NDSS Symposium
  2022}}. Internet Society.
\newblock
\urldef\tempurl%
\url{https://www.ndss-symposium.org/ndss-paper/auto-draft-210/}
\showURL{%
\tempurl}


\bibitem[Skolka et~al\mbox{.}(2019)]%
  {2019_anything_to_hide_studying_minified_and_obfuscated_code_in_the_web}
\bibfield{author}{\bibinfo{person}{Philippe Skolka},
  \bibinfo{person}{Cristian-Alexandru Staicu}, {and} \bibinfo{person}{Michael
  Pradel}.} \bibinfo{year}{2019}\natexlab{}.
\newblock \showarticletitle{Anything to Hide? Studying Minified and Obfuscated
  Code in the Web}. In \bibinfo{booktitle}{\emph{The World Wide Web Conf.}}
  \emph{(\bibinfo{series}{WWW '19})}. \bibinfo{publisher}{ACM},
  \bibinfo{address}{New York, NY, USA}, \bibinfo{pages}{1735--1746}.
\newblock
\showISBNx{9781450366748}
\href{https://doi.org/10.1145/3308558.3313752}{doi:\nolinkurl{10.1145/3308558.3313752}}


\bibitem[Sutherland et~al\mbox{.}(2006)]%
        {Sutherland2006}
\bibfield{author}{\bibinfo{person}{Iain Sutherland}, \bibinfo{person}{George~E.
  Kalb}, \bibinfo{person}{Andrew Blyth}, {and} \bibinfo{person}{Gaius Mulley}.}
  \bibinfo{year}{2006}\natexlab{}.
\newblock \showarticletitle{An empirical examination of the reverse engineering
  process for binary files}.
\newblock \bibinfo{journal}{\emph{Computers \& Security}} \bibinfo{volume}{25},
  \bibinfo{number}{3} (\bibinfo{year}{2006}), \bibinfo{pages}{221--228}.
\newblock


\bibitem[Tian et~al\mbox{.}(2022)]%
        {obfuscation_recognition}
\bibfield{author}{\bibinfo{person}{Zhenzhou Tian}, \bibinfo{person}{Hengchao
  Mao}, \bibinfo{person}{Yaqian Huang}, \bibinfo{person}{Jie Tian}, {and}
  \bibinfo{person}{Jinrui Li}.} \bibinfo{year}{2022}\natexlab{}.
\newblock \showarticletitle{Fine-Grained Obfuscation Scheme Recognition on
  Binary Code}. In \bibinfo{booktitle}{\emph{Digital Forensics and Cyber
  Crime}}. \bibinfo{publisher}{Springer Int'l Publishing},
  \bibinfo{address}{Cham}, \bibinfo{pages}{215--228}.
\newblock
\showISBNx{978-3-031-06365-7}


\bibitem[Tofighi-Shirazi et~al\mbox{.}(2019)]%
        {obfuscation_detection7}
\bibfield{author}{\bibinfo{person}{Ramtine Tofighi-Shirazi},
  \bibinfo{person}{Irina~M\u{a}riuca As\u{a}voae}, {and}
  \bibinfo{person}{Philippe Elbaz-Vincent}.} \bibinfo{year}{2019}\natexlab{}.
\newblock \showarticletitle{Fine-grained static detection of obfuscation
  transforms using ensemble-learning and semantic reasoning}. In
  \bibinfo{booktitle}{\emph{Proc.\ 9th Workshop on Software Security,
  Protection, and Reverse Engineering}} \emph{(\bibinfo{series}{SSPREW9 '19})}.
  \bibinfo{publisher}{ACM}, \bibinfo{address}{New York, NY, USA}, Article
  \bibinfo{articleno}{4}, \bibinfo{numpages}{12}~pages.
\newblock
\showISBNx{9781450377461}
\href{https://doi.org/10.1145/3371307.3371313}{doi:\nolinkurl{10.1145/3371307.3371313}}


\bibitem[Tsoutsos and Maniatakos(2017)]%
        {hash2}
\bibfield{author}{\bibinfo{person}{Nektarios~Georgios Tsoutsos} {and}
  \bibinfo{person}{Michail Maniatakos}.} \bibinfo{year}{2017}\natexlab{}.
\newblock \showarticletitle{Obfuscating branch decisions based on encrypted
  data using MISR and hash digests}. In \bibinfo{booktitle}{\emph{2017 Asian
  Hardware Oriented Security and Trust Symposium (AsianHOST)}}.
  \bibinfo{pages}{115--120}.
\newblock
\href{https://doi.org/10.1109/AsianHOST.2017.8354005}{doi:\nolinkurl{10.1109/AsianHOST.2017.8354005}}


\bibitem[Van~den Broeck et~al\mbox{.}(2021)]%
        {jens21}
\bibfield{author}{\bibinfo{person}{Jens Van~den Broeck}, \bibinfo{person}{Bart
  Coppens}, {and} \bibinfo{person}{Bjorn De~Sutter}.}
  \bibinfo{year}{2021}\natexlab{}.
\newblock \showarticletitle{Obfuscated integration of software protections}.
\newblock \bibinfo{journal}{\emph{Int'l Journal of Information Security}}
  \bibinfo{volume}{20} (\bibinfo{date}{02} \bibinfo{year}{2021}),
  \bibinfo{pages}{73--101}.
\newblock
\href{https://doi.org/10.1007/s10207-020-00494-8}{doi:\nolinkurl{10.1007/s10207-020-00494-8}}


\bibitem[Vasilescu et~al\mbox{.}(2017)]%
        {2017_recovering_clear_natural_identifiers_from_obfuscated_js_names}
\bibfield{author}{\bibinfo{person}{Bogdan Vasilescu}, \bibinfo{person}{Casey
  Casalnuovo}, {and} \bibinfo{person}{Premkumar Devanbu}.}
  \bibinfo{year}{2017}\natexlab{}.
\newblock \showarticletitle{Recovering Clear, Natural Identifiers from
  Obfuscated JS Names}. In \bibinfo{booktitle}{\emph{Proc.\ 11th Joint Meeting
  on Foundations of Software Engineering}} \emph{(\bibinfo{series}{ESEC/FSE
  2017})}. \bibinfo{publisher}{ACM}, \bibinfo{address}{New York, NY, USA},
  \bibinfo{pages}{683--693}.
\newblock
\showISBNx{9781450351058}
\href{https://doi.org/10.1145/3106237.3106289}{doi:\nolinkurl{10.1145/3106237.3106289}}


\bibitem[Viticchi\'{e} et~al\mbox{.}(2016)]%
        {viticchie2016reactive}
\bibfield{author}{\bibinfo{person}{Alessio Viticchi\'{e}},
  \bibinfo{person}{Cataldo Basile}, \bibinfo{person}{Andrea Avancini},
  \bibinfo{person}{Mariano Ceccato}, \bibinfo{person}{Bert Abrath}, {and}
  \bibinfo{person}{Bart Coppens}.} \bibinfo{year}{2016}\natexlab{}.
\newblock \showarticletitle{Reactive Attestation: Automatic Detection and
  Reaction to Software Tampering Attacks}. In \bibinfo{booktitle}{\emph{Proc.\
  ACM Workshop on Software PROtection}} \emph{(\bibinfo{series}{SPRO '16})}.
  \bibinfo{publisher}{ACM}, \bibinfo{pages}{73–84}.
\newblock
\showISBNx{9781450345767}


\bibitem[Votipka et~al\mbox{.}(2019)]%
        {Votipka2019}
\bibfield{author}{\bibinfo{person}{D. Votipka}, \bibinfo{person}{S. Rabin},
  \bibinfo{person}{K. Micinski}, \bibinfo{person}{J.~S. Foster}, {and}
  \bibinfo{person}{M.~L. Mazurek}.} \bibinfo{year}{2019}\natexlab{}.
\newblock \showarticletitle{An observational investigation of reverse
  engineers' process and mental models}. In \bibinfo{booktitle}{\emph{Extended
  Abstracts of the 2019 CHI Conf.\ on Human Factors in Computing Systems}}.
\newblock
\href{https://doi.org/10.1145/3290607.3313040}{doi:\nolinkurl{10.1145/3290607.3313040}}


\bibitem[WALA(2024)]%
        {WALA}
WALA \bibinfo{year}{2024}\natexlab{}.
\newblock \bibinfo{title}{{WAtson Libraries for Analysis (WALA)}}.
\newblock \bibinfo{howpublished}{\url{https://github.com/wala/WALA}}.
\newblock


\bibitem[Wang et~al\mbox{.}(2000)]%
        {flattening}
\bibfield{author}{\bibinfo{person}{Chenxi Wang}, \bibinfo{person}{Jonathan
  Hill}, \bibinfo{person}{John Knight}, {and} \bibinfo{person}{Jack Davidson}.}
  \bibinfo{year}{2000}\natexlab{}.
\newblock \bibinfo{booktitle}{\emph{Software Tamper Resistance: Obstructing
  Static Analysis of Programs}}.
\newblock \bibinfo{type}{{T}echnical {R}eport} CS-2000-12.
  \bibinfo{institution}{University of Virginia, Department of Computer
  Science}.
\newblock
\href{https://doi.org/10.18130/V36T9V}{doi:\nolinkurl{10.18130/V36T9V}}


\bibitem[Wang and Rountev(2017)]%
        {obfuscation_detection5}
\bibfield{author}{\bibinfo{person}{Yan Wang} {and} \bibinfo{person}{Atanas
  Rountev}.} \bibinfo{year}{2017}\natexlab{}.
\newblock \showarticletitle{Who Changed You? Obfuscator Identification for
  Android}. In \bibinfo{booktitle}{\emph{IEEE/ACM 4th Int'l Conf. on Mobile
  Software Engineering and Systems (MOBILESoft)}}. \bibinfo{pages}{154--164}.
\newblock
\href{https://doi.org/10.1109/MOBILESoft.2017.18}{doi:\nolinkurl{10.1109/MOBILESoft.2017.18}}


\bibitem[Wang and Shao(2003)]%
        {cognitive_functional_size}
\bibfield{author}{\bibinfo{person}{Yingxu Wang} {and} \bibinfo{person}{Jingqiu
  Shao}.} \bibinfo{year}{2003}\natexlab{}.
\newblock \showarticletitle{Measurement of the cognitive functional complexity
  of software}. In \bibinfo{booktitle}{\emph{Proc.\ The Second IEEE Int'l
  Conf.\ on Cognitive Informatics}}. \bibinfo{pages}{67--74}.
\newblock
\href{https://doi.org/10.1109/COGINF.2003.1225955}{doi:\nolinkurl{10.1109/COGINF.2003.1225955}}


\bibitem[Wohlin et~al\mbox{.}(2024)]%
        {Wohlin}
\bibfield{author}{\bibinfo{person}{C. Wohlin}, \bibinfo{person}{P. Runeson},
  \bibinfo{person}{M. H\"ost}, \bibinfo{person}{M.C. Ohlsson},
  \bibinfo{person}{B. Regnell}, {and} \bibinfo{person}{A. Wessl\'en}.}
  \bibinfo{year}{2024}\natexlab{}.
\newblock \bibinfo{booktitle}{\emph{Experimentation in Software Engineering}
  (\bibinfo{edition}{2nd} ed.)}.
\newblock \bibinfo{publisher}{Kluwer Academic Publishers}.
\newblock


\bibitem[Wong et~al\mbox{.}(2021)]%
        {wong2021inside}
\bibfield{author}{\bibinfo{person}{Yong~M. Wong}, \bibinfo{person}{M. Landen},
  \bibinfo{person}{M. Antonakakis}, \bibinfo{person}{D.~M. Blough},
  \bibinfo{person}{E.~M. Redmiles}, {and} \bibinfo{person}{M. Ahamad}.}
  \bibinfo{year}{2021}\natexlab{}.
\newblock \showarticletitle{An inside look into the practice of malware
  analysis}. In \bibinfo{booktitle}{\emph{Proc.\ ACM SIGSAC Conf.\ on Computer
  and Communications Security}}. \bibinfo{pages}{3053--3069}.
\newblock


\bibitem[Xie et~al\mbox{.}(2016)]%
        {hash1}
\bibfield{author}{\bibinfo{person}{Xin Xie}, \bibinfo{person}{Bin Lu},
  \bibinfo{person}{Daofu Gong}, \bibinfo{person}{Xiangyang Luo}, {and}
  \bibinfo{person}{Fenlin Liu}.} \bibinfo{year}{2016}\natexlab{}.
\newblock \showarticletitle{Random table and hash coding-based binary code
  obfuscation against stack trace analysis}.
\newblock \bibinfo{journal}{\emph{IET Information Security}}
  \bibinfo{volume}{10}, \bibinfo{number}{1} (\bibinfo{year}{2016}),
  \bibinfo{pages}{18--27}.
\newblock
\href{https://doi.org/10.1049/iet-ifs.2013.0137}{doi:\nolinkurl{10.1049/iet-ifs.2013.0137}}
\showeprint{https://ietresearch.onlinelibrary.wiley.com/doi/pdf/10.1049/iet-ifs.2013.0137}


\bibitem[Yadegari and Debray(2014)]%
        {bitleveltaint}
\bibfield{author}{\bibinfo{person}{Babak Yadegari} {and}
  \bibinfo{person}{Saumya Debray}.} \bibinfo{year}{2014}\natexlab{}.
\newblock \showarticletitle{Bit-{{Level Taint Analysis}}}. In
  \bibinfo{booktitle}{\emph{2014 {{IEEE}} 14th {{Int'l Working Conf.}} on
  {{Source Code Analysis}} and {{Manipulation}}}} (2014-09).
  \bibinfo{publisher}{{IEEE}}, \bibinfo{pages}{255--264}.
\newblock
\showISBNx{978-1-4799-6148-1}
\href{https://doi.org/10.1109/SCAM.2014.43}{doi:\nolinkurl{10.1109/SCAM.2014.43}}


\bibitem[Yadegari et~al\mbox{.}(2015a)]%
        {generic_deobfuscation}
\bibfield{author}{\bibinfo{person}{Babak Yadegari}, \bibinfo{person}{Brian
  Johannesmeyer}, \bibinfo{person}{Ben Whitely}, {and} \bibinfo{person}{Saumya
  Debray}.} \bibinfo{year}{2015}\natexlab{a}.
\newblock \showarticletitle{A Generic Approach to Automatic Deobfuscation of
  Executable Code}. In \bibinfo{booktitle}{\emph{IEEE Symposium on Security and
  Privacy}}. \bibinfo{pages}{674--691}.
\newblock
\href{https://doi.org/10.1109/SP.2015.47}{doi:\nolinkurl{10.1109/SP.2015.47}}


\bibitem[Yadegari et~al\mbox{.}(2015b)]%
        {2015_a_generic_approach_to_automatic_deobfuscation_of_executable_code}
\bibfield{author}{\bibinfo{person}{Babak Yadegari}, \bibinfo{person}{Brian
  Johannesmeyer}, \bibinfo{person}{Ben Whitely}, {and} \bibinfo{person}{Saumya
  Debray}.} \bibinfo{year}{2015}\natexlab{b}.
\newblock \showarticletitle{A generic approach to automatic deobfuscation of
  executable code}. In \bibinfo{booktitle}{\emph{IEEE Symposium on Security and
  Privacy}}. IEEE, \bibinfo{pages}{674--691}.
\newblock


\bibitem[Yong~Wong et~al\mbox{.}(2021)]%
        {practice_malware}
\bibfield{author}{\bibinfo{person}{Miuyin Yong~Wong}, \bibinfo{person}{Matthew
  Landen}, \bibinfo{person}{Manos Antonakakis}, \bibinfo{person}{Douglas~M.
  Blough}, \bibinfo{person}{Elissa~M. Redmiles}, {and}
  \bibinfo{person}{Mustaque Ahamad}.} \bibinfo{year}{2021}\natexlab{}.
\newblock \showarticletitle{An Inside Look into the Practice of Malware
  Analysis}. In \bibinfo{booktitle}{\emph{Proc.\ ACM SIGSAC Conf.\ on Computer
  and Communications Security}} \emph{(\bibinfo{series}{CCS '21})}.
  \bibinfo{publisher}{ACM}, \bibinfo{address}{New York, NY, USA},
  \bibinfo{pages}{3053–3069}.
\newblock
\showISBNx{9781450384544}
\href{https://doi.org/10.1145/3460120.3484759}{doi:\nolinkurl{10.1145/3460120.3484759}}


\bibitem[Zhang et~al\mbox{.}(2021)]%
        {obfuscation_detection}
\bibfield{author}{\bibinfo{person}{Xiaolu Zhang}, \bibinfo{person}{Frank
  Breitinger}, \bibinfo{person}{Engelbert Luechinger}, {and}
  \bibinfo{person}{Stephen O'Shaughnessy}.} \bibinfo{year}{2021}\natexlab{}.
\newblock \showarticletitle{Android application forensics: A survey of
  obfuscation, obfuscation detection and deobfuscation techniques and their
  impact on investigations}.
\newblock \bibinfo{journal}{\emph{Forensic Science Int'l: Digital
  Investigation}}  \bibinfo{volume}{39} (\bibinfo{year}{2021}),
  \bibinfo{pages}{301285}.
\newblock
\showISSN{2666-2817}
\href{https://doi.org/10.1016/j.fsidi.2021.301285}{doi:\nolinkurl{10.1016/j.fsidi.2021.301285}}


\bibitem[Zhao et~al\mbox{.}(2020)]%
        {obfuscation_detection8}
\bibfield{author}{\bibinfo{person}{Yujie Zhao}, \bibinfo{person}{Zhanyong
  Tang}, \bibinfo{person}{Guixin Ye}, \bibinfo{person}{Dongxu Peng},
  \bibinfo{person}{Dingyi Fang}, \bibinfo{person}{Xiaojiang Chen}, {and}
  \bibinfo{person}{Zheng Wang}.} \bibinfo{year}{2020}\natexlab{}.
\newblock \showarticletitle{Semantics-aware obfuscation scheme prediction for
  binary}.
\newblock \bibinfo{journal}{\emph{Computers \& Security}}  \bibinfo{volume}{99}
  (\bibinfo{year}{2020}), \bibinfo{pages}{102072}.
\newblock
\showISSN{0167-4048}
\href{https://doi.org/10.1016/j.cose.2020.102072}{doi:\nolinkurl{10.1016/j.cose.2020.102072}}


\bibitem[Zhou et~al\mbox{.}(2007)]%
        {mba}
\bibfield{author}{\bibinfo{person}{Yongxin Zhou}, \bibinfo{person}{Alec Main},
  \bibinfo{person}{Yuan~X. Gu}, {and} \bibinfo{person}{Harold Johnson}.}
  \bibinfo{year}{2007}\natexlab{}.
\newblock \showarticletitle{Information Hiding in Software with Mixed
  Boolean-Arithmetic Transforms}. In \bibinfo{booktitle}{\emph{Information
  Security Applications}}, Vol.~\bibinfo{volume}{4867}.
  \bibinfo{pages}{61--75}.
\newblock
\showISBNx{978-3-540-77535-5}
\href{https://doi.org/10.1007/978-3-540-77535-5_5}{doi:\nolinkurl{10.1007/978-3-540-77535-5_5}}


\end{thebibliography}



\appendix

\end{document}